\documentclass{article} 
\usepackage[final]{colm2026_conference}

\usepackage{microtype}
\usepackage{hyperref}
\usepackage{url}
\usepackage{booktabs}
\usepackage{enumitem}
\usepackage{graphicx}
\usepackage{subcaption}
\usepackage{threeparttable}
\usepackage{multirow}
\usepackage{makecell}
\usepackage[table]{xcolor}
\usepackage{algorithm}
\usepackage{algorithmic}
\usepackage{pifont}
\usepackage[utf8]{inputenc}
\usepackage{listings}
\definecolor{keywordcolor}{rgb}{0.7, 0.1, 0.1}
\definecolor{tacticcolor}{rgb}{0.0, 0.1, 0.6}
\definecolor{commentcolor}{rgb}{0.4, 0.4, 0.4}
\definecolor{symbolcolor}{rgb}{0.0, 0.1, 0.6}
\definecolor{sortcolor}{rgb}{0.1, 0.5, 0.1}
\definecolor{attributecolor}{rgb}{0.7, 0.1, 0.1}

\lstdefinelanguage{lean} {
mathescape=false,
texcl=false,
morekeywords=[1]{
import, prelude, protected, private, noncomputable, definition, meta, renaming,
hiding, parameter, parameters, begin, constant, constants,
lemma, variable, variables, theory,
print, theorem, example,
open, as, export, override, axiom, axioms, inductive, with,
structure, record, universe, universes,
alias, help, precedence, reserve, declare_trace, add_key_equivalence,
match, infix, infixl, infixr, notation, postfix, prefix, instance,
eval, reduce, check, end, this,
using, using_well_founded, namespace, section,
attribute, local, set_option, extends, include, omit, class,
raw, replacing,
calc, have, show, suffices, by, in, at, let, forall, Pi, fun,
exists, if, dif, then, else, assume, obtain, from, register_simp_ext, unless, break, continue,
mutual, do, def, run_cmd, const,
partial, mut, where, macro, syntax, deriving,
return, try, catch, for, macro_rules, declare_syntax_cat, abbrev, sorry},
morekeywords=[2]{Sort, Type, Prop},
morekeywords=[3]{
assumption,
apply, intro, intros, allGoals, 
generalize, clear, revert, done, exact,
refine, repeat, cases, rewrite, rw,
simp, simp_all, contradiction,
constructor, injection,
induction, 
prove_with, linarith, nlinarith, norm_num, contrapose, rfl, by_contra, push_neg, gcongr, aesop, field_simp, rfl, norm_cast, ring, ring_nf, positivity, omega, subst, rename_i, rcases, convert, ext, tauto, native_decide, use, by_cases,
to_theorem, operatorcount, quickcheck,
},
literate=
{α}{{\ensuremath{\mathrm{\alpha}}}}1
{β}{{\ensuremath{\mathrm{\beta}}}}1
{γ}{{\ensuremath{\mathrm{\gamma}}}}1
{δ}{{\ensuremath{\mathrm{\delta}}}}1
{ε}{{\ensuremath{\mathrm{\varepsilon}}}}1
{ζ}{{\ensuremath{\mathrm{\zeta}}}}1
{η}{{\ensuremath{\mathrm{\eta}}}}1
{θ}{{\ensuremath{\mathrm{\theta}}}}1
{ι}{{\ensuremath{\mathrm{\iota}}}}1
{κ}{{\ensuremath{\mathrm{\kappa}}}}1
{μ}{{\ensuremath{\mathrm{\mu}}}}1
{ν}{{\ensuremath{\mathrm{\nu}}}}1
{ξ}{{\ensuremath{\mathrm{\xi}}}}1
{π}{{\ensuremath{\mathrm{\mathnormal{\pi}}}}}1
{ρ}{{\ensuremath{\mathrm{\rho}}}}1
{σ}{{\ensuremath{\mathrm{\sigma}}}}1
{τ}{{\ensuremath{\mathrm{\tau}}}}1
{φ}{{\ensuremath{\mathrm{\varphi}}}}1
{χ}{{\ensuremath{\mathrm{\chi}}}}1
{ψ}{{\ensuremath{\mathrm{\psi}}}}1
{ω}{{\ensuremath{\mathrm{\omega}}}}1
{Γ}{{\ensuremath{\mathrm{\Gamma}}}}1
{Δ}{{\ensuremath{\mathrm{\Delta}}}}1
{Θ}{{\ensuremath{\mathrm{\Theta}}}}1
{Λ}{{\ensuremath{\mathrm{\Lambda}}}}1
{Σ}{{\ensuremath{\mathrm{\Sigma}}}}1
{Φ}{{\ensuremath{\mathrm{\Phi}}}}1
{Ξ}{{\ensuremath{\mathrm{\Xi}}}}1
{Ψ}{{\ensuremath{\mathrm{\Psi}}}}1
{Ω}{{\ensuremath{\mathrm{\Omega}}}}1
{ℵ}{{\ensuremath{\aleph}}}1
{≤}{{\ensuremath{\leq}}}1
{≥}{{\ensuremath{\geq}}}1
{≠}{{\ensuremath{\neq}}}1
{≈}{{\ensuremath{\approx}}}1
{≡}{{\ensuremath{\equiv}}}1
{≃}{{\ensuremath{\simeq}}}1
{≤}{{\ensuremath{\leq}}}1
{≥}{{\ensuremath{\geq}}}1
{∂}{{\ensuremath{\partial}}}1
{∆}{{\ensuremath{\triangle}}}1 
{∫}{{\ensuremath{\int}}}1
{∑}{{\ensuremath{\mathrm{\Sigma}}}}1
{Π}{{\ensuremath{\mathrm{\Pi}}}}1
{⊥}{{\ensuremath{\perp}}}1
{∞}{{\ensuremath{\infty}}}1
{∂}{{\ensuremath{\partial}}}1
{∓}{{\ensuremath{\mp}}}1
{±}{{\ensuremath{\pm}}}1
{×}{{\ensuremath{\times}}}1
{⊕}{{\ensuremath{\oplus}}}1
{⊗}{{\ensuremath{\otimes}}}1
{⊞}{{\ensuremath{\boxplus}}}1
{∇}{{\ensuremath{\nabla}}}1
{√}{{\ensuremath{\sqrt}}}1
{⬝}{{\ensuremath{\cdot}}}1
{•}{{\ensuremath{\cdot}}}1
{∘}{{\ensuremath{\circ}}}1
{⁻}{{\ensuremath{^{-}}}}1
{▸}{{\ensuremath{\blacktriangleright}}}1
{∧}{{\ensuremath{\wedge}}}1
{∨}{{\ensuremath{\vee}}}1
{¬}{{\ensuremath{\neg}}}1
{⊢}{{\ensuremath{\vdash}}}1
{⟨}{{\ensuremath{\langle}}}1
{⟩}{{\ensuremath{\rangle}}}1
{↦}{{\ensuremath{\mapsto}}}1
{←}{{\ensuremath{\leftarrow}}}1
{<-}{{\ensuremath{\leftarrow}}}1
{→}{{\ensuremath{\rightarrow}}}1
{↔}{{\ensuremath{\leftrightarrow}}}1
{⇒}{{\ensuremath{\Rightarrow}}}1
{⟹}{{\ensuremath{\Longrightarrow}}}1
{⇐}{{\ensuremath{\Leftarrow}}}1
{⟸}{{\ensuremath{\Longleftarrow}}}1
{∩}{{\ensuremath{\cap}}}1
{∪}{{\ensuremath{\cup}}}1
{⊂}{{\ensuremath{\subseteq}}}1
{⊆}{{\ensuremath{\subseteq}}}1
{⊄}{{\ensuremath{\nsubseteq}}}1
{⊈}{{\ensuremath{\nsubseteq}}}1
{⊃}{{\ensuremath{\supseteq}}}1
{⊇}{{\ensuremath{\supseteq}}}1
{⊅}{{\ensuremath{\nsupseteq}}}1
{⊉}{{\ensuremath{\nsupseteq}}}1
{∈}{{\ensuremath{\in}}}1
{∉}{{\ensuremath{\notin}}}1
{∋}{{\ensuremath{\ni}}}1
{∌}{{\ensuremath{\notni}}}1
{∅}{{\ensuremath{\emptyset}}}1
{∖}{{\ensuremath{\setminus}}}1
{†}{{\ensuremath{\dag}}}1
{ℕ}{{\ensuremath{\mathbb{N}}}}1
{ℤ}{{\ensuremath{\mathbb{Z}}}}1
{ℝ}{{\ensuremath{\mathbb{R}}}}1
{ℚ}{{\ensuremath{\mathbb{Q}}}}1
{ℂ}{{\ensuremath{\mathbb{C}}}}1
{⌞}{{\ensuremath{\llcorner}}}1
{⌟}{{\ensuremath{\lrcorner}}}1
{⦃}{{\ensuremath{\{\!|}}}1
{⦄}{{\ensuremath{|\!\}}}}1
{‖}{{\ensuremath{\|}}}1
{₁}{{\ensuremath{_1}}}1
{₂}{{\ensuremath{_2}}}1
{₃}{{\ensuremath{_3}}}1
{₄}{{\ensuremath{_4}}}1
{₅}{{\ensuremath{_5}}}1
{₆}{{\ensuremath{_6}}}1
{₇}{{\ensuremath{_7}}}1
{₈}{{\ensuremath{_8}}}1
{₉}{{\ensuremath{_9}}}1
{₀}{{\ensuremath{_0}}}1
{ᵢ}{{\ensuremath{_i}}}1
{ⱼ}{{\ensuremath{_j}}}1
{ₐ}{{\ensuremath{_a}}}1
{¹}{{\ensuremath{^1}}}1
{ₙ}{{\ensuremath{_n}}}1
{ₘ}{{\ensuremath{_m}}}1
{ₚ}{{\ensuremath{_p}}}1
{↑}{{\ensuremath{\uparrow}}}1
{↓}{{\ensuremath{\downarrow}}}1
{...}{{\ensuremath{\ldots}}}1
{·}{{\ensuremath{\cdot}}}1
{▸}{{\ensuremath{\triangleright}}}1
{∣}{{\ensuremath{\mid}}}1
{¹}{{\ensuremath{^1}}}1
{²}{{\ensuremath{^2}}}1
{³}{{\ensuremath{^3}}}1
{⁴}{{\ensuremath{^4}}}1
{⁵}{{\ensuremath{^5}}}1
{⁶}{{\ensuremath{^6}}}1
{⁷}{{\ensuremath{^7}}}1
{⁸}{{\ensuremath{^8}}}1
{⁹}{{\ensuremath{^9}}}1
{⁰}{{\ensuremath{^0}}}1
{Σ}{{\color{symbolcolor}\ensuremath{\Sigma}}}1
{Π}{{\color{symbolcolor}\ensuremath{\Pi}}}1
{∀}{{\color{symbolcolor}\ensuremath{\forall}}}1
{∃}{{\color{symbolcolor}\ensuremath{\exists}}}1
{λ}{{\color{symbolcolor}\ensuremath{\mathrm{\lambda}}}}1
{\$}{{\color{symbolcolor}\$}}1
{:=}{{\color{symbolcolor}:=}}1
{=}{{\color{symbolcolor}=}}1
{<|>}{{\color{symbolcolor}<|>}}1
{<\$>}{{\color{symbolcolor}<\$>}}1
{+}{{\color{symbolcolor}+}}1
{*}{{\color{symbolcolor}*}}1,
morecomment=[s][\color{commentcolor}]{/-}{-/},
morecomment=[l][\itshape \color{commentcolor}]{--},
showstringspaces=false,
keepspaces=true,
morestring=[b]",
morestring=[d],
tabsize=3,
extendedchars=false,
sensitive=true,
breaklines=true,
breakatwhitespace=true,
basicstyle=\ttfamily\small,
captionpos=b,
columns=[l]fullflexible,
identifierstyle={\ttfamily\color{black}},
keywordstyle=[1]{\ttfamily\color{keywordcolor}},
keywordstyle=[2]{\ttfamily\color{sortcolor}},
keywordstyle=[3]{\ttfamily\color{tacticcolor}},
keywordstyle=[4]{\ttfamily\color{attributecolor}},
stringstyle=\ttfamily,
commentstyle={\ttfamily\footnotesize },
}

\newcommand{\lean}[1]{\lstinline[language=lean,basicstyle=\ttfamily]!#1!}
\usepackage{tcolorbox}
\tcbuselibrary{listings,skins,breakable}
\definecolor{codebg}{HTML}{F5F5F5}
\definecolor{codeframe}{HTML}{CCCCCC}
\definecolor{cprimary}{HTML}{2166AC}
\definecolor{csecondary}{HTML}{B2182B}
\definecolor{ctertiary}{HTML}{2CA02C}
\tcbset{
  leanbase/.style={
    listing only,
    listing options={language=lean, numbers=none,
      basicstyle=\ttfamily\footnotesize, backgroundcolor=\color{white},
      aboveskip=0pt, belowskip=0pt},
    colback=white,
    boxrule=0pt,
    leftrule=2pt,
    arc=0pt,
    left=4pt, right=2pt, top=2pt, bottom=2pt,
  },
}
\newtcblisting{leancode}{
  leanbase,
  colframe=cprimary,
  before skip=0.8em,
  after skip=0.5em,
}
\newtcblisting{leancode-blue}{
  leanbase,
  colframe=cprimary,
  colback=cprimary!5,
  listing options={language=lean, numbers=none,
    basicstyle=\ttfamily\footnotesize, backgroundcolor=\color{cprimary!5},
    aboveskip=0pt, belowskip=0pt},
  before skip=2pt,
  after skip=2pt,
}
\newtcblisting{leancode-green}{
  leanbase,
  colframe=ctertiary,
  colback=ctertiary!5,
  listing options={language=lean, numbers=none,
    basicstyle=\ttfamily\footnotesize, backgroundcolor=\color{ctertiary!5},
    aboveskip=0pt, belowskip=0pt},
  before skip=2pt,
  after skip=2pt,
}
\newtcblisting{leancode-red}{
  leanbase,
  colframe=csecondary,
  colback=csecondary!5,
  listing options={language=lean, numbers=none,
    basicstyle=\ttfamily\footnotesize, backgroundcolor=\color{csecondary!5},
    aboveskip=0pt, belowskip=0pt},
  before skip=2pt,
  after skip=2pt,
}
\usepackage{tikz}
\usetikzlibrary{positioning, shapes.geometric, arrows.meta, calc, fit, backgrounds}

\usepackage{amsmath,amsfonts,bm}

\def\secref#1{section~\ref{#1}}

\def\eqref#1{equation~\ref{#1}}

\def\1{\bm{1}}

\DeclareMathAlphabet{\mathsfit}{\encodingdefault}{\sfdefault}{m}{sl}
\SetMathAlphabet{\mathsfit}{bold}{\encodingdefault}{\sfdefault}{bx}{n}

\usepackage{lineno}

\definecolor{darkblue}{rgb}{0, 0, 0.5}
\hypersetup{colorlinks=true, citecolor=darkblue, linkcolor=darkblue, urlcolor=darkblue}
\newcommand{\ours}{G\"{o}edel-Code-Prover-8B}
\NewDocumentCommand{\ziran}{ mO{} }{\textcolor{blue}{\textsuperscript{\textit{Ziran}}\textsf{\textbf{\small[#1]}}}}
\NewDocumentCommand{\haoyu}{ mO{} }{\textcolor{purple}{\textsuperscript{\textit{Haoyu}}\textsf{\textbf{\small[#1]}}}}
\NewDocumentCommand{\mike} { mO{} }{\textcolor{cyan}{\textsuperscript{\textit{Mike}}\textsf{\textbf{\small[#1]}}}}

\title{Goedel-Code-Prover: Hierarchical Proof Search for \\ Open State-of-the-Art Code Verification}

\author{Zenan Li$^{1,*}$, Ziran Yang$^{2,*}$, Deyuan He$^{3}$, Haoyu Zhao$^{3}$, Andrew Zhao$^{3}$, \\
\textbf{Shange Tang$^{2}$, Kaiyu Yang$^{4}$, Aarti Gupta$^{3}$, Zhendong Su$^{1}$, Chi Jin$^{2}$} \\
$^1$ETH Z\"urich, $^2$Princeton Language and Intelligence, \\
$^3$Department of Computer Science, Princeton University, $^4$Apodex \\
$^*$Equal contribution \\
\texttt{\{zenan.li@inf.ethz.ch, zirany@princeton.edu, chij@princeton.edu\}}
}

\begin{document}

\ifcolmsubmission
\linenumbers
\fi

\maketitle

\begin{abstract}
Large language models (LLMs) can generate plausible code but offer limited guarantees of correctness. 
Formally verifying that implementations satisfy specifications requires constructing machine-checkable proofs, a task that remains beyond current automation. We propose a hierarchical proof search framework for automated code verification in Lean~4 that decomposes complex verification goals into structurally simpler subgoals before attempting tactic-level proving. Central to our approach is a principled decomposition score that combines constructive justification with structural effectiveness. The same score serves as both the training reward and the inference-time ranking criterion, aligning optimization and deployment. We train \emph{\ours}, a single unified policy for both decomposition and completion, through supervised initialization followed by hybrid reinforcement learning, where a continuous decomposition reward supports planning exploration while supervised replay stabilizes proof generation. On three Lean-based code verification benchmarks comprising 427 tasks, our 8B-parameter model achieves a 62.0\% prove success rate, a 2.6$\times$ improvement over the strongest baseline under the reported inference settings.
We further observe consistent inference-time scaling: success rates improve monotonically with search iterations and sampling budget, while whole-proof baselines plateau within the evaluated budgets.
\end{abstract}

\section{Introduction}
\label{sec:intro}

Large language models (LLMs) have rapidly advanced code generation, achieving strong performance on programming benchmarks and increasingly serving as real-world coding assistants~\citep{chen2021evaluating,li2022competition,alharbi2026automatic,li2026advances}.
Yet while LLM-based coding agents routinely generate unit tests and reproduction scripts to validate their outputs, empirical studies show that such testing cannot certify the absence of subtle logical errors, boundary-case violations, or specification mismatches~\citep{jesse2023large,liu2023your,al2026model}. 
As LLMs are deployed in safety-critical environments, bridging the gap between ``code that appears to work'' and ``code that is provably correct'' becomes increasingly urgent.

Formal verification offers a principled solution by expressing specifications (see Figure~\ref{fig:running-example} for an example) in a formal language and constructing machine-checked proofs of correctness.
Lean~4~\citep{moura2021lean4}, a modern interactive theorem prover, provides a powerful platform for such verification tasks, with a growing ecosystem of libraries (e.g., \lean{Mathlib}~\citep{mathlib2020community} and \lean{CSLib}~\citep{barrett2026cslib}).
Figure~\ref{fig:running-example} illustrates a typical verification task: given a function \lean{FindSingleNumber} that returns the unique element from a list, the specification encodes a precondition and a postcondition as Lean predicates, and the top-level theorem asserts that the implementation satisfies this specification.
The \lean{sorry} placeholder marks the proof obligation to be discharged.
Even for this seemingly simple problem, constructing a complete proof requires discovering auxiliary lemmas about list filtering, membership, and inductive reasoning over the input structure, none of which are suggested by the specification itself.
Writing such proofs by hand remains labor-intensive and demands substantial expertise, limiting widespread adoption of formal verification.

\begin{figure}[t]
\begin{leancode-blue}
-- Precondition: every element appears 1 or 2 times; exactly one is unique
def FindSingleNumber_precond (nums : List Int) : Prop :=
  let numsCount := nums.map (fun x => nums.count x)
  numsCount.all (fun c => c = 1 ∨ c = 2) ∧ numsCount.count 1 = 1
\end{leancode-blue}
\begin{leancode-green}
-- Implementation: find the unique element that appears exactly once
def filterlist (x : Int) (nums : List Int) : List Int :=
  nums.filter (fun y => y == x)

def FindSingleNumber (nums : List Int)
    (h : FindSingleNumber_precond nums) : Int :=
  match nums.filter (fun x => (filterlist x nums).length == 1) with
  | x :: _ => x
  | []     => 0
\end{leancode-green}
\begin{leancode-blue}
-- Postcondition: result is the unique element; others appear twice
def FindSingleNumber_postcond (nums : List Int) (result : Int)
    (h : FindSingleNumber_precond nums) : Prop :=
  nums.length > 0 ∧ (filterlist result nums).length = 1 ∧
  ∀ (x : Int), x ∈ nums → x = result ∨ (filterlist x nums).length = 2
\end{leancode-blue}
\begin{leancode-red}
-- Verification goal (proof to be synthesized automatically)
theorem FindSingleNumber_spec (nums : List Int)
    (h : FindSingleNumber_precond nums) :
    FindSingleNumber_postcond nums (FindSingleNumber nums h) h := by
  sorry
\end{leancode-red}
\caption{A code verification task in Lean~4. {\color{cprimary}\textbf{Blue}}: specification (precondition and postcondition). {\color{ctertiary}\textbf{Green}}: implementation. {\color{csecondary}\textbf{Red}}: verification goal (\lean{sorry} marks the proof obligation). Full proof in Appendix~\ref{app:running-example}.}
\label{fig:running-example}
\end{figure}

Recent work has explored leveraging LLMs to automate proof synthesis in interactive theorem provers (ITPs) such as Lean~4 and Isabelle~\citep{li2024survey,yang2024formal}.
State-of-the-art systems have achieved remarkable success in formal mathematics, reaching olympiad-level performance~\citep{ren2025deepseek, lin2025goedel, xin2025scaling, hubert2025olympiad, wang2025kimina}.
These methods typically operate in a \emph{decompose-and-prove} paradigm: the model first decomposes a complex theorem into simpler intermediate lemmas (or subgoals), then proves each lemma individually using tactic generation.

Despite this progress, these methods do not transfer directly to code verification due to two fundamental mismatches.
The first is \emph{ungrounded decomposition}: unlike mathematics, where proof reasoning pervades textbooks and online discussions, virtually no natural-language corpus describes how program specifications are proved~\citep{myers2004art,ammann2017introduction}. LLMs therefore lack the informal-to-formal bridge that enables effective decomposition in mathematics, and their proposed subgoals are frequently imprecise or no simpler than the original goal.
The second is \emph{compound domain shift}: each program introduces its own functions and invariants without library support (\emph{concept proliferation}), and discharging subgoals requires program-specific tactics that differ substantially from those prevalent in mathematics (\emph{tactic distribution shift}).

\begin{figure*}[t]
\centering
\resizebox{\textwidth}{!}{%
\begin{tikzpicture}[
    >=latex,
    rootnode/.style={
        rectangle, rounded corners=3pt, draw=black!60, very thick,
        fill=blue!10, minimum height=0.8cm,
        font=\footnotesize, align=center, inner sep=5pt
    },
    subnode/.style={
        rectangle, rounded corners=3pt, draw=black!50, thick,
        fill=blue!6, minimum height=0.7cm,
        font=\scriptsize, align=center, inner sep=4pt
    },
    leafnode/.style={
        rectangle, rounded corners=3pt, draw=black!50, thick,
        fill=green!8, minimum height=0.7cm,
        font=\scriptsize, align=center, inner sep=4pt
    },
    provednode/.style={
        rectangle, rounded corners=3pt, draw=green!55!black, very thick,
        fill=green!18, minimum height=0.65cm,
        font=\scriptsize\bfseries, align=center, inner sep=4pt
    },
    opcount/.style={
        circle, draw=purple!55!black, thick, fill=purple!12,
        font=\tiny\bfseries, inner sep=1pt, minimum size=0.38cm
    },
    scorelabel/.style={
        rectangle, rounded corners=2pt, draw=black!35, thick,
        fill=yellow!8, font=\scriptsize, inner sep=3pt, align=center
    },
    chkbox/.style={
        rectangle, rounded corners=2pt, draw=black!40, thick,
        fill=gray!6, font=\scriptsize, inner sep=3pt, align=left
    },
    policynode/.style={
        rectangle, rounded corners=3pt, draw=orange!55!black, thick,
        fill=orange!10, minimum height=0.5cm,
        font=\scriptsize, align=center, inner sep=3pt
    },
    leannode/.style={
        rectangle, rounded corners=3pt, draw=teal!55!black, thick,
        fill=teal!8, minimum height=0.65cm,
        font=\footnotesize, align=center, inner sep=5pt
    },
    arr/.style={->, thick, draw=black!45},
    darr/.style={->, draw=black!30, densely dashed},
    treeline/.style={-, thick, draw=black!30},
    note/.style={font=\tiny, text=black!70},
    slabel/.style={font=\footnotesize\bfseries},
]


\node[rootnode] (root) at (0, 0) {%
    Theorem $G$: \;\texttt{FindSingleNumber} satisfies specification};
\node[opcount, right=3pt of root] (d0) {18};
\node[right=3pt of d0, font=\scriptsize, text=purple!55!black] {operator footprint $d(\cdot)$};

\node[policynode] (pi1) at (0, -0.95) {LLM $\pi$};
\draw[arr, draw=black!40] (root.south) -- (pi1.north);

\node[subnode] (l1) at (-3.8, -2.0) {%
    $L_1$: result is the\\[-1pt]unique element};
\node[opcount, right=2pt of l1] (d1) {7};

\node[subnode] (l2) at (3.8, -2.0) {%
    $L_2$: other elements\\[-1pt]appear exactly twice};
\node[opcount, right=2pt of l2] (d2) {8};

\draw[treeline] (pi1.south) -- ++(0, -0.2) -| (l1.north);
\draw[treeline] (pi1.south) -- ++(0, -0.2) -| (l2.north);

\node[chkbox] (chk) at (0, -2.0) {%
    $\checkmark$ proof reconstruction\\[-1pt]
    $\checkmark$ quickcheck filter\\[-1pt]
    $\checkmark$ operator footprint $\downarrow$};

\node[policynode] (pi2) at (-3.8, -2.9) {LLM $\pi$};
\draw[arr, draw=black!40] (l1.south) -- (pi2.north);

\node[leafnode] (l11) at (-6.2, -4.0) {%
    $L_{1.1}$: filterlist length\\[-1pt]equals List.count};
\node[opcount, right=2pt of l11] (d11) {4};

\node[leafnode] (l12) at (-3.0, -4.0) {%
    $L_{1.2}$: some element\\[-1pt]has count 1};
\node[opcount, right=2pt of l12] (d12) {4};

\node[leafnode] (l13) at (0.2, -4.0) {%
    $L_{1.3}$: scanner finds\\[-1pt]the unique element};
\node[opcount, right=2pt of l13] (d13) {5};

\draw[treeline] (pi2.south) -- ++(0, -0.2) -| (l11.north);
\draw[treeline] (pi2.south) -- ++(0, -0.2) -| (l12.north);
\draw[treeline] (pi2.south) -- ++(0, -0.2) -| (l13.north);

\node[policynode] (pi3) at (3.8, -2.9) {LLM $\pi$};
\draw[arr, draw=black!40] (l2.south) -- (pi3.north);

\node[leafnode] (l21) at (3.8, -4.0) {%
    $L_{2.1}$: count-1 element\\[-1pt]is unique};
\node[opcount, right=2pt of l21] (d21) {4};

\draw[treeline] (pi3.south) -- (l21.north);

\node[scorelabel] (s1) at (7.8, -2.0) {$\begin{aligned}S &= \textit{validity} \cdot \textit{d-reduction}\\[-1pt] &= 1.0 \times 0.54\end{aligned}$};
\node[scorelabel] (s2) at (7.8, -4.0) {$S = v \cdot r = 1.0 \times 0.33$};

\coordinate (box-left) at (-7.6, 0);
\coordinate (box-right) at (9.4, 0);
\coordinate (box-top) at (0, 1.3);
\coordinate (box-mid) at (0, -4.85);
\coordinate (box-bot) at (0, -7.4);

\begin{scope}[on background layer]
\fill[blue!4] (box-left |- box-top) rectangle (box-right |- box-mid);
\end{scope}
\draw[black!30, thick] (box-left |- box-top) rectangle (box-right |- box-mid);
\node[slabel, text=blue!45!black, anchor=north west, xshift=3pt, yshift=-2pt] at (box-left |- box-top) {%
    Stage 1: Recursive Lemma Decomposition};


\node[leafnode, minimum width=1.1cm] (fl1) at (-5.0, -6.15) {$L_{1.1}$};
\node[leafnode, minimum width=1.1cm] (fl2) at (-3.6, -6.15) {$L_{1.2}$};
\node[leafnode, minimum width=1.1cm] (fl3) at (-2.2, -6.15) {$L_{1.3}$};
\node[leafnode, minimum width=1.1cm] (fl4) at (-0.8, -6.15) {$L_{2.1}$};

\node[policynode, minimum width=2.4cm, minimum height=0.65cm, font=\footnotesize] (policy) at (2.0, -6.15) {LLM $\pi$};

\node[leannode, minimum width=2.2cm] (lean) at (5.2, -6.15) {Lean Kernel};

\node[provednode, minimum width=1.8cm] (proved) at (7.8, -6.15) {Verified $\checkmark$};

\node[draw=black!35, dashed, rounded corners=3pt,
    fit=(fl1)(fl2)(fl3)(fl4), inner xsep=5pt, inner ysep=5pt] (leafbox) {};
\node[note, below=1pt of leafbox, font=\tiny] {leaf lemmas};

\draw[arr] (leafbox) -- (policy);
\draw[arr] (policy) -- node[above, font=\tiny] {tactics} (lean);
\draw[arr, draw=green!45!black] (lean) -- node[above, font=\tiny] {pass} (proved);

\draw[darr, draw=teal!40!black, rounded corners=3pt]
    (lean.south) -- ++(0, -0.4) -| node[below, pos=0.25, font=\tiny, text=black!40] {error feedback} (policy.south);

\begin{scope}[on background layer]
\fill[green!4] (box-left |- box-mid) rectangle (box-right |- box-bot);
\end{scope}
\draw[black!30, thick] (box-left |- box-mid) rectangle (box-right |- box-bot);
\node[slabel, text=green!40!black, anchor=north west, xshift=3pt, yshift=-2pt] at (box-left |- box-mid) {%
    Stage 2: Lemma Completion};

\node[scorelabel, minimum width=14.0cm] (dualscore) at (1.0, -7.85) {%
    Decomposition Score: \;$S = v(\{L_i\}; G) \cdot r(\{L_i\}; G)$
    \quad$\longrightarrow$\quad
    training reward \;\&\; inference-time ranking};

\end{tikzpicture}%
}
\caption{
\textbf{Overview.}
\emph{Stage~1:} The goal is recursively decomposed into simpler sub-lemmas; the score $S$ ranks candidates by validity and structural reduction in operator footprint $d(\cdot)$.
\emph{Stage~2:} Leaf lemmas are proved via iterative tactic generation with Lean~4 compiler feedback. The score serves as both training reward and inference-time ranking criterion.
}
\label{fig:framework}
\end{figure*}

In this work, we propose a \emph{hierarchical proof search} framework for automated code verification in Lean~4, in which the verification goal is recursively decomposed into a series of lemmas that are then proved iteratively (Figure~\ref{fig:framework}).
To achieve this, we train \emph{\ours}, a single unified policy that performs both decomposition and completion.
Since the quality of decomposition is the primary determinant of success, we introduce a \emph{decomposition score} as both the training reward and inference-time search criterion, evaluating candidates along two axes: constructive justification (whether proposed subgoals provably entail the parent theorem and survive automated counterexample search via \emph{quickcheck}~\citep{claessen2000quickcheck, lean2024plausible}), and structural simplicity (measured by a goal complexity metric).
After supervised initialization, we apply hybrid reinforcement learning as a modest refinement: the continuous score optimizes decomposition while supervised replay stabilizes completion.

At inference time, the model first decomposes the verification goal into subgoals (\emph{lemma decomposition}), then proves each via iterative tactic generation (\emph{lemma completion}); the decomposition score ranks candidate decompositions, while the unified policy performs both stages.

We evaluate our framework on three Lean~4 code verification benchmarks: Verina~\citep{ye2025verina}, Clever~\citep{thakur2025clever}, and AlgoVeri~\citep{zhao2026algoveri}, comprising 427 tasks in total.
Our 8B-parameter model achieves success rates of 68.8\%, 54.0\%, and 62.3\% respectively, a 2.6$\times$ improvement over the strongest baseline and a 38.2-point absolute gain over the best neural prover under the reported inference settings.
Verified proofs average 8--17 auxiliary lemmas and 130--167 lines of proof code across benchmarks, with the most complex reaching 680 lines, demonstrating that our framework can sustain deep structured reasoning over non-trivial verification tasks.
We further observe consistent inference-time scaling: success rates improve monotonically with search iterations and sampling budget, indicating unsaturated scaling potential.

In summary, our contributions are:
(1)~a hierarchical proof search framework that decomposes complex verification goals into simpler subgoals before tactic-level proving;
(2)~a decomposition score combining constructive justification with structural simplicity, used as both training reward and inference-time search criterion;
(3)~a hybrid training pipeline that jointly trains decomposition and completion within a unified policy, with RL providing a modest but consistent refinement over supervised initialization;
and (4)~extensive evaluation showing that a purpose-trained 8B model outperforms the reported frontier-model and neural-prover baselines under their respective inference settings.

\section{Lean-based Code Verification}
\label{sec:problem}

We model automated code verification in Lean~4 using a Hoare-style specification framework.
Concretely, we are given:

\begin{itemize}[leftmargin=*,nosep]
\item a \emph{program} $C$ (a Lean definition or function),
\item a \emph{precondition} $P$ and a \emph{postcondition} $Q$, both expressed as Lean predicates.
\end{itemize}

The verification objective is to establish a Hoare-style triple
\begin{equation*}
\{P\}\; C \;\{Q\},
\end{equation*}
which asserts that for any input $x$ satisfying $P(x)$, the evaluation of $C(x)$ yields a result $v$ such that $Q(x, v)$ holds. 
In the functional setting of Lean, $C$ is typically a pure function, and the triple is encoded as a Lean proposition relating inputs and outputs of $C$.
In our running example (Figure~\ref{fig:running-example}), $C = \texttt{FindSingleNumber}$, $P$ requires every element appears once or twice with exactly one unique element, and $Q$ asserts the output is that unique element.

Formally, the input to the verification task is a tuple $(C, P, Q)$ expressed in Lean, and the goal is to construct a machine-checkable proof term $\tau$ such that the Lean kernel accepts $\tau$ as a valid derivation of the verification goal $G$:
\begin{equation*}
G \;:\; \forall x, P(x) \to Q(x, C(x)).
\end{equation*}%
We assume that the program $C$ and its associated predicates $P$ and $Q$ are already formalized in Lean; translating informal specifications into Lean is out of scope.
The core challenge is therefore the synthesis of the proof term $\tau$, which often involves quantifier elimination, complex induction, invariant generation, and the formulation of auxiliary assertions.

When we employ an LLM to automate this synthesis, the task reduces to interactive tactic generation: the LLM observes a \emph{goal state} (local hypotheses plus target proposition), issues a tactic command, and the Lean kernel checks it, updates the proof state, and generates new subgoals. This loop continues until all goals are discharged or the attempt is abandoned.

We choose Lean~4 over SMT-based verifiers (e.g., Dafny, Verus) because its interactive proof protocol exposes explicit goal states and fine-grained compiler diagnostics, enabling iterative LLM-driven search; a detailed comparison is provided in Appendix~\ref{app:why-lean}.
As discussed in Section~\ref{sec:intro}, automated code verification poses three key challenges beyond formal mathematics---ungrounded decomposition, concept proliferation, and tactic distribution shift---which together create a compound search problem with an exceptionally fragile search space (see Appendix~\ref{app:challenges} for an extended analysis).

\section{Hierarchical Verification Framework}
\label{sec:framework}

Our framework centers on a \emph{decomposition score} that quantifies candidate lemma decompositions and serves consistently as both the training reward and the inference-time ranking criterion.
We first define the score (\secref{subsec:score}) and then describe the unified training and inference pipeline (\secref{sec:learning_inference}).

\subsection{Score for Lemma Decomposition}
\label{subsec:score}

Lemma decomposition is a critical bottleneck: a good decomposition turns a global correctness theorem into provable subgoals, whereas a poor one produces invalid or equally intractable obligations.
We seek a \emph{computable} score that correlates with downstream provability, is cheap to evaluate online, and serves consistently as both the training reward and the inference-time ranking criterion.
We define the score along two complementary axes: \emph{(i)~constructive justification} and \emph{(ii)~decomposition effectiveness}.

{\bf Constructive justification}. 
A lemma decomposition is only meaningful if the proposed lemmas ($L_1,\dots,L_k$) logically entail the parent theorem ($G$). 
Specifically, a candidate decomposition must provide a proof reconstruction: a proof term $\pi_{\text{parent}}$, produced by the LLM as part of its decomposition output, such that Lean can verify the implication $(L_1 \wedge \dots \wedge L_k) \Rightarrow G$.
We require that $\pi_{\text{parent}}$ explicitly invokes each proposed lemma, ensuring the decomposition is not merely a syntactic list but a constructively justified reduction of the original goal.

However, constructive justification alone does not guarantee the validity of the subgoals: a false lemma (e.g., \lean{l = []} $\wedge$ \lean{l.length > 0}) can technically prove the parent theorem but shifts the burden to an impossible sub-task.
To filter out such cases, we integrate \emph{quickcheck}~as a semantic gate: it randomly samples concrete inputs and checks whether each proposed lemma holds. If a counterexample is found for any $L_i$, the entire decomposition is discarded before expensive proof search is attempted.

{\bf Decomposition effectiveness.}
A decomposition is effective if it reduces the difficulty of the original verification task.
In Hoare-triple terms, consider splitting $\{P\}\,C\,\{Q\}$ into $\{P\}\,C_1\,\{R\}$ and $\{R\}\,C_2\,\{Q\}$, where $C_1; C_2$ is a partition of the program $C$ and $R$ is an intermediate assertion characterizing the state between the two fragments.
Such a decomposition is productive when each resulting sub-goal is strictly simpler than the original, either by simplifying the logical assertions (i.e., $R$ simpler than $P$ and $Q$) or the programs (i.e., $C_1, C_2$ simpler than $C$).

To quantify this notion of simplicity, we observe that in Lean's abstract syntax tree (AST), both assertions and programs are uniformly expressed through operators: logical operators (connectives and quantifiers) and program operators (arithmetic, bitwise, or data-structure constructors).
We define the structural difficulty $d(G)$ of a verification goal $G$ as its \emph{operator footprint}: the total number of logical and domain-specific operator occurrences in its AST.
This metric captures reasoning cost because each operator corresponds to a specific class of proof obligations: reducing logical operators simplifies propositional reasoning, and reducing computational operators simplifies reasoning about program behavior.

We illustrate this with the running example (Figure~\ref{fig:running-example}). The goal \lean{FindSingleNumber_spec} combines specification operators ($\forall$, $\to$, $\wedge$, $\vee$, $\in$, \lean{=}) with program operators (\lean{filterlist}, \lean{List.count}, \lean{List.map}, \lean{List.length}), yielding $d(G)=18$.
A decomposition splits $G$ into $L_1$: ``\lean{filterlist} on the unique element returns a singleton list'' ($d(L_1)=7$) and $L_2$: ``every other element appears in \lean{filterlist} exactly twice'' ($d(L_2)=8$).
Each sub-lemma isolates a single property of \lean{filterlist} with a smaller footprint, confirming a positive structural reduction.

\subsection{Unified Policy Learning and Inference}
\label{sec:learning_inference}

We implement two custom Lean tactics to compute the score: \lean{operatorcount} for the operator footprint and \lean{quickcheck} for automated counterexample search.
The score $S$ combines a binary validity gate with a continuous structural reduction ratio.
A candidate decomposition is valid only if (i) the proposed lemmas collectively discharge $G$ via proof reconstruction, and (ii) every lemma survives quickcheck without counterexamples.
We encode this as
\begin{equation*}
v(L_1, \dots, L_k; G) = \mathbb{1}_{\text{proof}}(L_1, \dots, L_k; G) \cdot \prod_{i=1}^{k} \mathbb{1}_{\text{qc}}(L_i),
\end{equation*}
where $\mathbb{1}_{\text{proof}} (L_1, \dots, L_k; G) = 1$ if proof reconstruction succeeds and $0$ otherwise, and $\mathbb{1}_{\text{qc}}(L_i) = 1$ if no counterexample is found for $L_i$ and $0$ otherwise.

Next, we quantify how much simpler the sub-goals are relative to the original.
We aggregate the individual footprints $d(L_i)$ via LogSumExp, which approximates $\max_i d(L_i)$ while remaining sensitive to reductions in any individual subgoal:
\begin{equation*}
\bar{d}(L_1, \dots, L_k) = T \cdot \log \sum_{i=1}^{k} \exp\left(d(L_i)/T\right),
\end{equation*}
where $T$ controls the smoothness.
The structural reduction ratio is then
\begin{equation*}
r(L_1,\dots,L_k; G) = \max(1 - \frac{\bar{d}(L_1, \dots, L_k)}{d(G)},\, 0).
\end{equation*}

The final decomposition score is $S = r(L_1,\dots,L_k; G) \cdot v(L_1, \dots, L_k; G)$.
Because this score is used identically during training (as the reward) and inference (as the ranking criterion), policy learning and tree-search control are optimized against exactly the same objective.
For clarity, the diagnostic plots in Figures~\ref{fig:verina-correlation}, \ref{fig:train-correlation}, \ref{fig:score-dist-before}, and~\ref{fig:score-dist-after} retain the unnormalized score prior to its conversion into $S$ (their axes are labeled ``Score'').
This raw quantity is oriented as residual difficulty---lower values indicate simpler decompositions---and should not be confused with the normalized, higher-is-better score $S$ defined above.

\subsubsection{Training Pipeline}
We train the same autoregressive policy for both planning and proving in two stages: (1) supervised initialization with scaffolded data; (2) hybrid reinforcement learning that optimizes decomposition while stabilizing completion.

{\bf Supervised fine-tuning (SFT)}.
Using frontier models (GPT-5.2 and Gemini-3-Flash), we generate scaffolded trajectories for both lemma decomposition and lemma completion.
This process yields 281K decomposition input–output pairs and 151K completion pairs.
We merge the two types of trajectories and fine-tune a Qwen-3-8B backbone to obtain an initial policy $\pi_0$.
At this stage, the model already supports both decomposition and completion behaviors under the same prompting scaffolding.

{\bf Hybrid reinforcement learning (RL)}.
A central challenge in jointly optimizing decomposition and completion is reward mismatch: decomposition receives a dense, continuous score in $[0,1]$, while completion yields only a sparse binary signal (proof accepted or rejected).
Naively merging these objectives causes decomposition gradients to dominate and completion proficiency to stagnate.

We resolve this by decoupling the learning objectives while maintaining a shared parameter space.
Decomposition is optimized via group relative policy optimization (GRPO), utilizing the structural score $S$ to drive exploration and curriculum expansion.
Completion is stabilized through supervised replay of high-quality proof trajectories (generated by either the policy itself or frontier models), avoiding reliance on brittle binary signals.
Algorithm~\ref{alg:hybrid-rl} (Appendix~\ref{app:algorithms}) summarizes the full loop.
For completion, we adopt a \emph{policy-first} strategy: $\pi$ attempts each lemma first, and a frontier model is invoked only when $\pi$ fails within $m$ attempts, ensuring that $\pi$ is exposed to its own successful completions for replay.
Additionally, successfully decomposed lemmas are fed back into the problem set, providing an ever-expanding curriculum of progressively simpler subgoals.

\subsubsection{Inference Pipeline}

Inference proceeds in two sequential stages: the decomposition stage runs first, iteratively breaking down the original goal until a budget is exhausted, after which the completion stage takes over and attempts to prove each remaining leaf lemma.
We use the term \emph{iteration} to refer to a single decomposition or completion step within its respective stage.

{\bf Lemma Decomposition}.
At each decomposition iteration, the policy selects the open goal with the highest operator footprint as the decomposition target, concentrating search effort on the most complex remaining obligation.
The policy then rolls out a decomposition for the selected goal; the result is verified via proof reconstruction and quickcheck, and discarded if either check fails.
Upon acceptance, the original goal is replaced by its sub-lemmas in the goal set, and operator footprints are computed for the new goals.
This process repeats until the decomposition budget is exhausted or the open-lemma cap is reached.

{\bf Lemma Completion}.
After the decomposition stage concludes, the remaining open lemmas are passed to the completion stage.
At each completion iteration, the policy generates a candidate proof for each unclosed lemma and submits it to the Lean compiler.
If compilation fails, the error diagnostics are fed back to the policy, which revises the proof accordingly.
This generate-and-refine loop repeats until all proofs are accepted or the budget is exhausted.

Algorithm~\ref{alg:inference} (Appendix~\ref{app:algorithms}) summarizes the full inference pipeline.
Beyond increasing the per-run budget, performance can be further improved by launching $k$ independent runs in parallel and accepting the first successful proof (pass@$k$).

\section{Experiments}
\label{sec:experiments}

\subsection{Experimental Setup}
\label{subsec:setup}

{\bf Benchmarks.}
We evaluate on three Lean~4 code verification benchmarks.
Verina~\citep{ye2025verina} contains 189 tasks from introductory programming exercises, covering code generation, specification synthesis, and proof generation.
Clever~\citep{thakur2025clever} provides 161 problems with formal specifications designed to avoid test-case leakage and trivial solutions.
AlgoVeri~\citep{zhao2026algoveri} comprises 77 classical algorithms with identical functional contracts across Dafny, Verus, and Lean; we adopt its Lean subset.
Among these, Verina ships complete programs with preconditions, implementations, and postconditions, forming a self-contained evaluation suite.
Clever and AlgoVeri only release specifications without reference implementations; we therefore prompt GPT-5.2 to generate the code and manually verify the results.
Together, the three benchmarks yield 427 verification tasks spanning straightforward functional correctness to deep algorithmic reasoning.

{\bf Baselines.}
We compare against two categories of baselines.
(1)~\emph{Frontier reasoning models}: Claude-Opus-4.6, Gemini-3-Flash, GPT-5.2-Pro, GPT-5.3-Codex, and DeepSeek-V3.2-Speciale, each prompted to produce a complete proof in a single pass.
(2)~\emph{Neural provers}: Kimina-Prover-72B~\citep{wang2025kimina}, DeepSeek-Prover-V2-671B~\citep{ren2025deepseek}, Goedel-Prover-V2-32B~\citep{lin2025goedel}, and BFS-Prover-V2-32B~\citep{xin2025scaling}, which are trained on large-scale formal proof corpora; among them, BFS-Prover employs best-first tree search over tactic sequences, while the others generate whole proofs.
All baselines use default inference configurations and report pass@128.
These comparisons are not compute-matched: our method performs iterative hierarchical search, whereas most baselines sample independent whole proofs or tactic sequences.
We therefore report outcomes under the stated inference settings without attributing differences to model size alone; Appendix~\ref{app:budget-comparison} details the available budgets.

{\bf Implementation details.}
We fine-tune a Qwen-3-8B backbone~\citep{yang2025qwen3} following our training pipeline; we refer to the resulting policy as \emph{\ours}.
SFT uses 281K decomposition and 151K completion trajectories, followed by hybrid RL.
At inference, the decomposition stage runs up to 128 iterations or until adding another decomposition would exceed 32 open lemmas; the completion stage refines each lemma for up to 128 iterations.
Additional training hyperparameters, prompt templates, and inference configurations are provided in Appendix~\ref{app:implementation}.

{\bf Training data and decontamination.}
The 5K training problems are title/description-disjoint from all evaluation tasks.
Audits on boilerplate-stripped Lean code find no shared substring of at least 200 characters between any benchmark and training implementation; an embedding audit flags only 2/427 pairs, both from Verina, and manual inspection finds that neither admits proof transfer.
Threshold-level results and the flagged pairs are reported in Appendix~\ref{app:decontamination}.

{\bf Metrics.}
We report the \emph{prove success rate}: the fraction of problems for which a complete, Lean-verified proof is produced within the inference budget.
To ensure soundness, we only admit the standard axioms \lean{propext}, \lean{Classical.choice}, and \lean{Quot.sound}; proofs invoking \lean{Lean.ofReduceBool} or \lean{Lean.trustCompiler} are rejected, as these bypass the kernel's deductive checking and could mask unsound reasoning.

\subsection{Overall Effectiveness}
\label{subsec:rq1}

\begin{figure}[t]
    \centering
    \includegraphics[width=\linewidth]{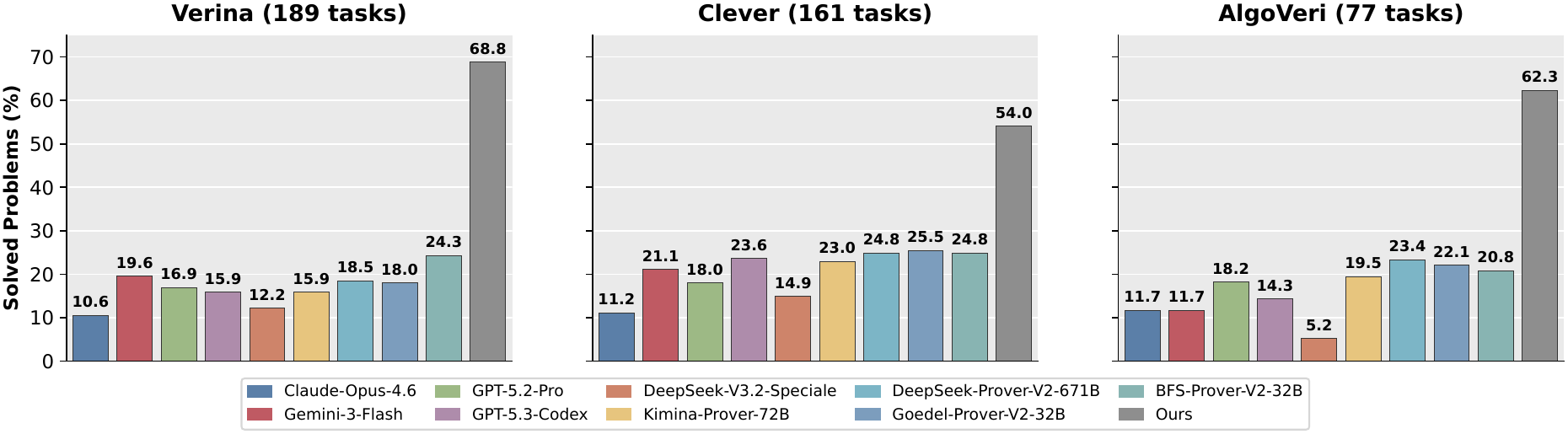}
    \caption{Percentage of solved problems by baselines and our framework across three benchmarks.
    Baselines are evaluated with parallel generation under a Pass@128 budget;
    our method operates under a search-based inference setting using \ours.
    Our framework outperforms all baselines by a substantial margin across every benchmark.}
    \label{fig:results}
\end{figure}

As shown in Figure~\ref{fig:results}, under the reported inference settings our framework achieves prove success rates of 68.8\%, 54.0\%, and 62.3\% on Verina, Clever, and AlgoVeri, respectively, yielding an overall rate of 62.0\% across all 427 tasks.
In addition, quickcheck disproves 23, 10, and 14 problems on Verina, Clever, and AlgoVeri by producing concrete counterexamples.\footnote{Quickcheck is also applied directly to the top-level goal. When a counterexample is found, it is provided to the LLM, which then uses it as evidence to disprove the problem.}
Among frontier reasoning models, the strongest (GPT-5.3-Codex) reaches only 23.6\% on Clever and below 20\% on Verina and AlgoVeri, confirming that single-pass proof generation is insufficient for non-trivial verification.
Among neural provers, BFS-Prover-V2-32B achieves the best overall score of 23.8\%, while our framework is 38.2 percentage points higher under the reported, non-compute-matched settings.

On average, each verified problem requires 8 to 17 auxiliary lemmas and 130 to 167 lines of proof code across benchmarks, with the longest proofs reaching 680, 559, and 534 lines on Verina, Clever, and AlgoVeri (detailed statistics in Table~\ref{tab:lemma-proof-stats}, Appendix~\ref{sec:appendix}).

\subsection{Inference-Time Scaling}
\label{subsec:rq2}

Our framework exposes two orthogonal scaling axes: the \emph{per-run search budget} (number of iterations within each decomposition or completion stage) and the \emph{parallel sampling budget} (pass@$k$, where $k$ independent runs are launched and the first successful proof is accepted).
As shown in Figure~\ref{fig:reduction-rate} (Appendix~\ref{app:reduction-rate}), the residual operator-footprint ratio decreases steadily as iterations progress, indicating that the framework progressively identifies more effective goal simplifications with additional compute.
Increasing the pass@$k$ budget further lowers this ratio across all three benchmarks, confirming that broader parallel search at the decomposition stage produces structurally simpler sub-goals.

\begin{figure}[t]
    \centering
    \includegraphics[width=\linewidth]{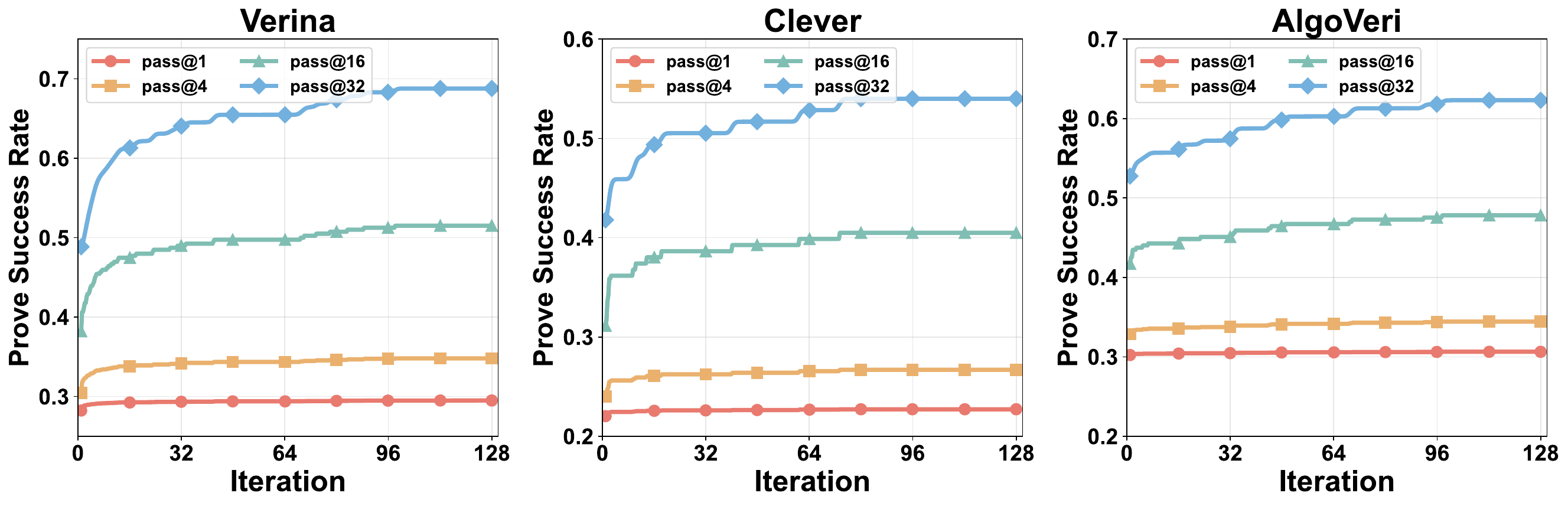}
    \caption{Prove success rate vs.\ completion iterations under different pass@$k$ budgets.}
    \label{fig:prove-success-rate}
\end{figure}

Figure~\ref{fig:prove-success-rate} shows that the prove success rate grows steadily with both more completion iterations and larger pass@$k$, demonstrating clear inference-time scaling across all benchmarks.
The persistent gap between pass@1 and pass@32 suggests that the framework benefits substantially from parallel search and has not yet saturated its scaling potential within the evaluated budget.
By contrast, the pass@$k$ curves of baseline verifiers plateau rapidly (see Appendix~\ref{app:pass-at-k}), indicating that brute-force sampling without hierarchical decomposition yields diminishing returns.

\subsection{Decomposition Analysis}
\label{subsec:rq3}

\begin{figure}[t]
    \centering
    \includegraphics[width=\linewidth]{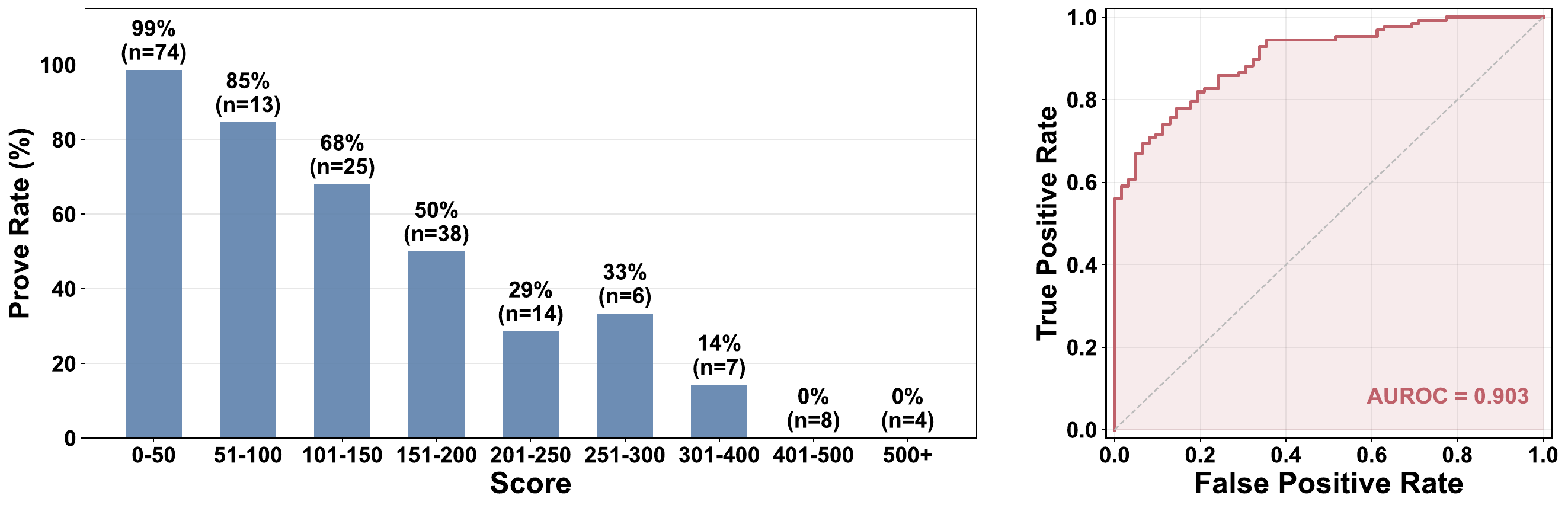}
    \caption{Unnormalized decomposition score vs.\ prove rate on Verina. Lower raw scores indicate simpler decompositions and separate instances proved within the evaluated budget from those left unsolved (AUROC = 0.903).}
    \label{fig:verina-correlation}
\end{figure}

\begin{table}[t]
\begin{minipage}[t]{0.42\linewidth}
    \centering
    \captionof{table}{Decomposition failure rates. Proof Failed: \% of iterations where proof reconstruction rejects the decomposition. QC Failed: \% of runs where at least one lemma fails quickcheck.}
    \label{tab:decompose-error}
    \setlength{\tabcolsep}{6pt}
    \renewcommand{\arraystretch}{1.2}
    \begin{tabular}{@{}l|rr@{}}
    \toprule
    Dataset & Proof Failed & QC Failed \\
    \midrule
    Verina    & 59.4 & 46.4 \\
    Clever    & 44.9 & 31.8 \\
    AlgoVeri  & 52.8 & 32.6 \\
    \bottomrule
    \end{tabular}
\end{minipage}
\hfill
\begin{minipage}[t]{0.55\linewidth}
    \centering
    \captionof{table}{Ablation study on each component. We progressively replace each component with our trained model to isolate its contribution.}
    \label{tab:ablation}
    \begin{threeparttable}
    \setlength{\tabcolsep}{8pt}
    \renewcommand{\arraystretch}{1.2}
    \begin{tabular}{ll|c}
    \toprule
    \textbf{Decomposition} & \textbf{Completion} & Verina \\
    \midrule
    ---           & Gemini-3-Flash  & 19.6 \\
    GPT-5.2-Pro          & Gemini-3-Flash  & 54.4 \\
    Ours          & Gemini-3-Flash  & 58.2 \\
    GPT-5.2-Pro           & Ours    & 59.2 \\
    \midrule
    Ours          & Ours    & \textbf{68.8} \\
    \bottomrule
    \end{tabular}
    \end{threeparttable}
\end{minipage}
\end{table}

Table~\ref{tab:decompose-error} quantifies how each criterion filters out unproductive decompositions.
Among 32 parallel runs per problem, 31.8\%--46.4\% are discarded entirely because at least one proposed lemma fails quickcheck (QC Failed).
Separately, proof reconstruction rejects 44.9\%--59.4\% of individual decomposition iterations (Proof Failed), confirming that constructing a logically sound reduction remains a major bottleneck.
Because these rates use different denominators (runs versus iterations), they should not be added directly; together, they characterize complementary failure modes filtered by quickcheck and proof reconstruction.

Figure~\ref{fig:verina-correlation} further shows that the unnormalized decomposition score is a strong inverse predictor of downstream provability: on Verina, lower raw scores correlate with higher prove rates, achieving an AUROC of 0.903.
Because the plotted quantity is the pre-normalization form of $S$, this result supports using the normalized, higher-is-better $S$ to rank candidate decompositions during training and inference.
The same inverse association appears on our training data (Appendix~\ref{app:train-correlation}).

\subsection{Ablation Study}
\label{subsec:rq4}

\begin{table}[h]
    \centering
    \caption{Matched-budget comparison of the SFT-only policy and the policy after RL on Verina. Entries are prove success rates (\%).}
    \label{tab:sft-only}
    \setlength{\tabcolsep}{6pt}
    \renewcommand{\arraystretch}{1.15}
    \begin{tabular}{lrrrr}
    \toprule
    Model & Pass@1 & Pass@10 & Pass@20 & Pass@32 \\
    \midrule
    SFT only & 26.9 & 44.9 & 53.9 & 66.1 \\
    SFT + RL & 29.1 & 46.0 & 57.1 & 68.8 \\
    \midrule
    Difference & +2.2 & +1.1 & +3.2 & +2.7 \\
    \bottomrule
    \end{tabular}
\end{table}

Table~\ref{tab:ablation} isolates the contributions of our trained decomposition and completion modules by swapping each with baseline alternatives on Verina.
Without decomposition, directly prompting Gemini-3-Flash for whole-proof generation achieves only 19.6\%, confirming that hierarchical decomposition is essential.
Adding GPT-5.2-Pro decomposition raises the rate to 54.4\%; replacing it with our trained decomposer yields 58.2\%, showing the value of domain-specific decomposition training.
Similarly, replacing Gemini-3-Flash completion with our trained completer (GPT-5.2-Pro + Ours: 59.2\%) shows that domain-specific completion training is at least as valuable as improved decomposition.
Combining both trained components achieves 68.8\%, surpassing every mixed configuration and confirming synergistic gains from joint training: the decomposer and completer co-adapt to each other's strengths.
Because these module swaps compare models trained with both SFT and RL, they do not isolate the RL contribution.
Table~\ref{tab:sft-only} therefore compares the same hierarchical policy before and after RL at matched inference budgets.
RL adds a smaller but consistent 1.1--3.2 percentage points; the bulk of performance comes from hierarchical search and supervised training, and the single training run does not establish statistical significance.
Separately, applying our framework to an off-the-shelf GPT-OSS-120B model raises Verina success from 20.1\% with whole-proof generation to 44.9\% with hierarchical search (Appendix~\ref{app:open-model}), showing that the structural benefit is not restricted to the frontier models used for data collection.

We also visualize how training shifts the unnormalized decomposition-score distribution in Appendix~\ref{app:score-distribution}: after training, raw scores shift toward lower values across all benchmarks, corresponding to simpler decompositions and higher normalized scores.

\section{Related work}
\label{sec:related}

LLM-based theorem proving has advanced in mathematics, from tactic-level generation~\citep{polu2020generative} and retrieval-augmented proving~\citep{yang2024leandojo} to large-scale neural provers~\citep{xin2024deepseek, wang2025kimina, lin2025goedel} and RL-driven hierarchical decomposition~\citep{hubert2025olympiad,dong2024proDRL}.
These advances do not transfer directly to code verification because of concept proliferation and tactic distribution shift (Appendix~\ref{app:challenges}).
For code verification, LLMs have generated SMT annotations~\citep{kamath2023finding, pei2023can}, produced or retrieved flat proofs~\citep{first2023baldur, thompson2024rango}, and driven tactic-level Isabelle search with symbolic repair~\citep{he2026neurosymbolic}.
The most closely related system, Quarry, ranks Rocq decompositions by predicted hammer solvability~\citep{zhang2026planning}; we instead train a unified Lean~4 policy using a code-specific decomposition score shared by training and inference.
Learned compositional abstractions have also been studied for inductive program synthesis~\citep{ellis2020dreamcoder}, but not for decomposing ITP proof obligations.
Our framework introduces a hierarchical planning layer with a continuous decomposition score, addressing the reward sparsity that limits prior RL approaches~\citep{xin2025scaling, lightman2024let}.
A detailed discussion is provided in Appendix~\ref{app:related}.

\section{Conclusion}
\label{sec:conclusion}

We presented a hierarchical proof search framework for code verification in Lean~4, centered on a decomposition score that serves as both the training reward and the inference-time selection criterion.
On three benchmarks, our 8B-parameter model achieves a 62.0\% overall prove success rate, a 2.6$\times$ improvement over the strongest baseline under the reported inference settings.
Its larger iterative inference budget is not compute-matched to parallel single-pass baselines; despite unsaturated scaling, reducing inference cost remains important.
Resource constraints limited RL to one run, precluding across-seed variance and significance estimates. We also lack comparable token, API-cost, and GPU-hour accounting and per-task cross-system AlgoVeri outcomes.
Further limitations are pre-formalized programs, syntactic operator counting, and single-procedure scope; informal specifications, semantic metrics, and interprocedural reasoning remain future work.

\newpage
\section*{Acknowledgments}
We thank the anonymous reviewers for their valuable comments. This work benefited from joint discussions and iterative refinement by all authors. Aarti Gupta acknowledges support from a Princeton SEAS Innovation Research Grant. Chi Jin acknowledges support from NSF grants OAC-2411299 and IIS-2239297 and the Princeton AI Lab Seed Fund.

\bibliography{colm2026_conference}
\bibliographystyle{colm2026_conference}

\appendix
\newpage
\section{Disclosure of LLM Usage}
\label{app:llm-disclosure}
As required by the COLM 2026 call for papers, we disclose all uses of large language models in preparing this work.
In the research pipeline, GPT-5.2 and Gemini-3-Flash were used to generate Lean formalizations and training trajectories as described in Section~\ref{sec:framework} and Appendix~\ref{app:implementation}; GPT-OSS-120B was used in the open-model experiment, and \texttt{text-embedding-3-large} was used in the decontamination audit.
LLM-assisted coding tools (Cursor and Claude Code) were used during software development, including experiment scripts and data-processing pipelines.
ChatGPT and Gemini were additionally used for language editing and for drafting camera-ready revisions from author-provided technical content.
The authors designed the research and experiments, verified generated formalizations and proofs using Lean and the checks described in the paper, analyzed the results, and take full responsibility for all final scientific claims.

\section{Extended Related Work}
\label{app:related}

\paragraph{LLM-based theorem proving and the math-code gap.}
Language models have achieved remarkable progress in formal theorem proving, predominantly in mathematics.
Starting from tactic-level generation~\citep{polu2020generative} and retrieval-augmented proving~\citep{yang2024leandojo}, the field has advanced through tree-search methods such as HyperTree Proof Search~\citep{lample2022hypertree} and, more recently, large-scale neural provers including DeepSeek-Prover~\citep{xin2024deepseek}, Kimina-Prover~\citep{wang2025kimina}, BFS-Prover~\citep{xin2025bfs}, and Goedel-Prover~\citep{lin2025goedel}.
Among these, Goedel-Prover is particularly noteworthy for its curriculum-driven synthetic data pipeline, which iteratively generates and verifies proof data to bootstrap prover capability.

However, the success of these systems does not directly transfer to code verification.
In mathematics, a well-curated library such as \lean{Mathlib}~\citep{mathlib2020community} provides a dense, reusable inventory of lemmas, and the space of core concepts is comparatively concentrated.
Code verification, by contrast, lacks a comparable lemma ecosystem: each program introduces its own function definitions, data-type invariants, and control-flow structures, effectively defining a fresh set of domain-specific concepts per task.
This \emph{concept proliferation} renders the proof distribution far more heterogeneous and precludes reliance on a fixed library of reusable lemmas, creating a substantial generalization gap for methods trained primarily on mathematical corpora.

\paragraph{Automated code verification with LLMs.}
Traditional program verification relies on SMT-based auto-active tools such as Dafny~\citep{leino2010dafny} and Verus~\citep{lattuada2023verus}, where correctness is established by discharging verification conditions through automated solvers.
Recent efforts have explored using LLMs to assist these workflows, for example by generating loop invariants or annotations~\citep{kamath2023finding, pei2023can}.
In the Lean ecosystem, Baldur~\citep{first2023baldur} demonstrated whole-proof generation for software verification, and Rango~\citep{thompson2024rango} introduced adaptive retrieval-augmented proving.
However, these approaches either operate in a flat, single-pass generation mode or rely on tactic-level tree search without explicit goal decomposition.
Our framework differs by introducing a planning layer that decomposes verification goals into structurally simpler sub-lemmas before invoking tactic-level proving, enabling the system to tackle problems whose complexity exceeds what flat generation can handle.

\paragraph{Hierarchical proof search and reinforcement learning.}
Decomposing complex goals into simpler sub-problems has a long history in automated reasoning~\citep{li2024survey}.
In neural theorem proving, Draft-Sketch-Prove~\citep{jiang2022draft} generates informal proof sketches before formalizing them, and subgoal-based approaches~\citep{achim2025aristotle, ren2025deepseek} have explored decomposition for mathematical theorems.
AlphaProof~\citep{hubert2025olympiad} achieved IMO-gold-level performance by combining reinforcement learning with a planning-and-proving paradigm.
The most closely related prior system, ProD-RL~\citep{dong2024proDRL}, trains an Isabelle/AFP prover to propose and prove intermediate lemmas, rewarding correct and novel lemmas discovered during training.
Our setting and mechanism differ: we target Lean~4 code verification and use a continuous score that combines proof reconstruction, quickcheck filtering, and operator-footprint reduction, with the same score also ranking decompositions at inference time.
DreamCoder~\citep{ellis2020dreamcoder} instead learns reusable symbolic library abstractions for inductive program synthesis and does not decompose proof obligations through an interactive theorem prover.
Our work adapts this paradigm to code verification, where useful decompositions differ fundamentally: intermediate assertions must capture semantic program properties (such as inductive invariants and termination measures) rather than syntactic refinements of the goal statement.
On the reinforcement learning side, prior work has applied expert iteration~\citep{polu2020generative}, off-policy RL with tree search~\citep{xin2025scaling}, and process reward models~\citep{lightman2024let} to improve proof search.
A persistent challenge is reward sparsity: binary proof success/failure signals provide limited gradient information.
Our hybrid RL approach addresses this by introducing a continuous decomposition score as a dense reward for the planning stage, while stabilizing the proving stage through supervised replay.

\section{Verification Paradigms and Why Lean}
\label{app:why-lean}

Modern program verification predominantly relies on SMT-based ``auto-active'' verifiers such as Dafny~\citep{leino2010dafny} and Verus~\citep{lattuada2023verus}.
These tools translate high-level code and annotations into verification conditions (VCs) discharged by automated solvers such as CVC5~\citep{barbosa2022cvc5} or Z3~\citep{de2008z3}.
While highly effective for systems engineering, the reasoning process is largely mediated by external SMT solvers, which act as black boxes from the LLM's perspective.
When a VC fails, the solver typically returns a generic counterexample or timeout rather than a structured diagnostic, making it difficult for an LLM agent to identify the root cause and iteratively refine its annotations.

We adopt Lean~4~\citep{moura2021lean4} as our verification platform for LLM-driven proof automation.
From a foundational standpoint, Lean offers two structural advantages well-suited to LLM-driven workflows:
(i) \emph{Small trusted kernel.} Lean relies on a minimal logical kernel as its sole verification authority, so even when an LLM produces unpredictable or creative proof attempts, correctness is guaranteed by a small, well-audited checker rather than a complex solver stack.
(ii) \emph{Rich library ecosystem.} Lean's dependent type theory supports higher-order and richly structured specifications, and libraries such as \lean{Mathlib}~\citep{mathlib2020community} and \lean{CSLib}~\citep{barrett2026cslib} provide an extensive knowledge base of verified lemmas and tactics that an LLM can directly invoke, facilitating the encoding of properties that extend beyond first-order SMT reasoning.

More importantly, Lean exposes the entire proof construction process to the LLM agent.
Unlike opaque VC failures in SMT-based systems, Lean provides explicit goal states, local hypotheses, and fine-grained compiler diagnostics at each proof step.
This transparency enables the LLM to plan proof strategies by inspecting open goals, execute tactics with immediate feedback, and debug failed attempts using structured error messages.
Such an interactive loop is particularly suitable for iterative, model-driven verification.

Additionally, recent cross-system evaluations enable a controlled outcome comparison.
The \lean{AlgoVeri} benchmark~\citep{zhao2026algoveri} translates the same set of algorithmic problems into Dafny, Verus, and Lean under aligned functional contracts, enabling a controlled comparison across verification paradigms.
On its 77 aligned-contract tasks, the published Gemini-3-Flash evaluation reports success rates of 40.3\% for Dafny, 24.7\% for Verus, and 7.8\% for Lean, whereas our Lean framework reaches 62.3\% on the same task set.
This controls the functional contracts but not the end-to-end inference compute, so the 22.0-point difference from Dafny should be interpreted as an outcome comparison rather than a compute-matched systems comparison.
The result nevertheless shows that iterative proof interaction can make Lean competitive despite forgoing the raw automation of SMT solvers.

\section{Key Challenges in Code Verification}
\label{app:challenges}

Despite recent advances in LLM-based theorem proving, automated code verification remains fundamentally harder than its mathematical counterpart.
State-of-the-art planning-and-proving methods have achieved remarkable success in formal mathematics, reaching competition- and IMO-level performance~\citep{hubert2025olympiad}, yet their efficacy does not transfer to code verification.
We identify three key challenges that explain this gap.

{\bf Ungrounded decomposition.}
Complex verification goals are seldom provable in a single pass; the prover must carefully plan its proof strategy and decompose the goal into auxiliary lemmas whose proofs collectively entail the top-level statement.
In formal mathematics, LLMs acquire strong planning ability from extensive pretraining over textbooks, competition solutions, and online discussions, enabling them to propose meaningful decompositions even for challenging problems~\citep{hubert2025olympiad, chen2025seed}.
However, virtually no natural-language corpus describes how a program's specification is proved; reasoning about program correctness has no comparable textual footprint in LLM pretraining data.
Without this grounding, frontier models frequently produce subgoals that are semantically invalid or no simpler than the original objective.

{\bf Concept proliferation.}
Even given a correct decomposition, proving each subgoal remains difficult because each program introduces a fresh conceptual universe.
In formal mathematics, the relevant objects (real numbers, groups, combinatorial structures) form a relatively closed vocabulary supported by shared libraries such as Mathlib, and proofs can reuse a rich collection of existing lemmas.
In contrast, every function definition in a program brings new computational behavior, control flow patterns, and data-structure invariants whose properties must be established from scratch without prior library support~\citep{pei2023can, kamath2023finding}.
For example, the postcondition in Figure~\ref{fig:running-example} relies on \lean{filterlist}, a helper function defined within the program itself; no existing library contains lemmas about its behavior, so all necessary properties must be proved from scratch during verification.

{\bf Tactic distribution shift.}
Beyond the conceptual gap, the tactics needed to discharge code verification subgoals differ markedly from those prevalent in formal mathematics.
Mathematical proofs rely heavily on algebraic normalization tactics such as \lean{linarith}, \lean{polyrith}, and \lean{gcongr} over continuous domains.
Code verification, by contrast, operates over discrete and recursive domains (e.g., machine integers, bit-vectors, lists, and graphs), requiring specialized automation such as \lean{grind}, domain-specific decision procedures, and explicit recursive reasoning.
For instance, proving properties about \lean{filterlist} in our running example requires list induction and element-membership reasoning, handled by induction, simplification, or \lean{grind} rather than the algebraic tactics that dominate mathematical proofs.
Models trained primarily on mathematical corpora often fail to invoke these program-specific tools correctly, struggling to close even well-formulated subgoals.

{\bf The compound search challenge.}
Together, these three challenges create a compound search problem with an exceptionally fragile search space: a well-crafted set of subgoals can collapse a complex verification condition into tractable obligations, while a slightly mis-specified lemma renders the entire proof intractable.
This sensitivity motivates our approach, which trains a domain-specific decomposer guided by a principled scoring mechanism to identify decompositions that are both semantically valid and tactically provable.

\section{Complete Proof for the Running Example}
\label{app:running-example}

Below is the complete machine-checked proof for \texttt{FindSingleNumber\_spec} (the running example from Figure~\ref{fig:running-example}), as synthesized by our framework.
The implementation shown here is the full version used in the actual proof; Figure~\ref{fig:running-example} presents an equivalent but abbreviated form for readability.
The proof decomposes the top-level theorem into six auxiliary lemmas covering list filtering, element counting, uniqueness, and the recursive scanner, totaling over 130 lines of proof code.
Each block below corresponds to one lemma or the final theorem.

\newtcblisting{leancode-gray}{
  leanbase,
  colframe=codeframe,
  before skip=0.5em,
  after skip=0.5em,
  breakable,
}

\begin{leancode-gray}
-- Precondition and postcondition (specification)
def FindSingleNumber_precond (nums : List Int) : Prop :=
  let numsCount := nums.map (fun x => nums.count x)
  numsCount.all (fun c => c = 1 ∨ c = 2)
    ∧ numsCount.count 1 = 1

def FindSingleNumber_postcond (nums : List Int)
    (result : Int)
    (h : FindSingleNumber_precond nums) : Prop :=
  nums.length > 0
    ∧ (filterlist result nums).length = 1
    ∧ ∀ (x : Int), x ∈ nums →
        x = result ∨ (filterlist x nums).length = 2
\end{leancode-gray}

\begin{leancode-gray}
-- Helper function used in the specification
def filterlist (x : Int) (nums : List Int) : List Int :=
  let rec aux (lst : List Int) : List Int :=
    match lst with
    | []      => []
    | y :: ys => if y = x then y :: aux ys else aux ys
  aux nums

-- Implementation: find the unique element that appears exactly once
def FindSingleNumber (nums : List Int)
    (h_precond : FindSingleNumber_precond nums) : Int :=
  let rec findUnique (remaining : List Int) : Int :=
    match remaining with
    | [] => 0
    | x :: xs =>
      if (filterlist x nums).length = 1 then x
      else findUnique xs
  findUnique nums
\end{leancode-gray}

\begin{leancode-gray}
-- Lemma 1: filterlist length equals List.count
lemma length_filterlist_eq_count (x : Int) (nums : List Int) :
    (filterlist x nums).length = nums.count x := by
  classical
  have aux_eq : ∀ lst : List Int,
      filterlist.aux x lst = lst.filter (fun y => y = x) := by
    intro lst; induction lst with
    | nil => simp [filterlist.aux]
    | cons y ys ih =>
        by_cases h : y = x <;> simp [filterlist.aux, ih, h]
  have eq_filter :
      filterlist x nums = nums.filter (fun y => y = x) :=
    by simp [filterlist, aux_eq]
  simpa [eq_filter] using
    (List.count_eq_length_filter (l := nums) (a := x)).symm
\end{leancode-gray}

\begin{leancode-gray}
-- Lemma 2: the precondition guarantees a unique element exists
lemma exists_count_eq_one (nums : List Int)
    (h_precond : FindSingleNumber_precond nums) :
    ∃ x ∈ nums, nums.count x = 1 := by
  classical
  unfold FindSingleNumber_precond at h_precond
  rcases h_precond with ⟨_hall, huniq⟩
  have hmem : 1 ∈ (nums.map (fun x => nums.count x)) := by
    by_contra hnot
    have : (nums.map (fun x => nums.count x)).count 1 = 0 :=
      by simp [List.count_eq_zero, hnot]
    exact Nat.one_ne_zero (by simpa [this] using huniq)
  rcases List.mem_map.mp hmem with ⟨x, hx, hxcount⟩
  exact ⟨x, hx, by simpa using hxcount⟩
\end{leancode-gray}

\begin{leancode-gray}
-- Lemma 3: the recursive scanner returns a unique element
lemma findUnique_returns_length_one (nums : List Int) :
    ∀ remaining : List Int,
      (∃ u ∈ remaining, (filterlist u nums).length = 1) →
        (filterlist (FindSingleNumber.findUnique
          (nums := nums) remaining) nums).length = 1 := by
  classical
  intro remaining; induction remaining with
  | nil => intro hex; rcases hex with ⟨u, hu, _⟩; cases hu
  | cons x xs ih =>
      intro hex; simp [FindSingleNumber.findUnique]
      by_cases hx1 : (filterlist x nums).length = 1
      · simp [hx1]
      · have hex' :
            ∃ u ∈ xs, (filterlist u nums).length = 1 := by
          rcases hex with ⟨u, hu_mem, hu_len⟩; simp at hu_mem
          cases hu_mem with
          | inl hu_eq =>
              subst hu_eq; exact False.elim (hx1 hu_len)
          | inr hu_mem_xs =>
              exact ⟨u, hu_mem_xs, hu_len⟩
        simpa [hx1] using ih hex'
\end{leancode-gray}

\begin{leancode-gray}
-- Lemma 4: the algorithm returns a value with count = 1
lemma FindSingleNumber_result_characterization
    (nums : List Int)
    (h_precond : FindSingleNumber_precond nums) :
    nums.count (FindSingleNumber nums h_precond) = 1 := by
  classical
  rcases exists_count_eq_one nums h_precond
    with ⟨u, hu_mem, hu_count⟩
  have hu_len : (filterlist u nums).length = 1 :=
    by simp [length_filterlist_eq_count, hu_count]
  have hlen := by simpa [FindSingleNumber] using
    findUnique_returns_length_one nums nums
      ⟨u, hu_mem, hu_len⟩
  simpa [length_filterlist_eq_count] using hlen
\end{leancode-gray}

\begin{leancode-gray}
-- Lemma 5: uniqueness of the element with count = 1
lemma precond_unique_count_one (nums : List Int)
    (h_precond : FindSingleNumber_precond nums)
    {a b : Int} (ha : a ∈ nums) (hb : b ∈ nums)
    (hca : nums.count a = 1) (hcb : nums.count b = 1) :
    a = b := by
  classical
  unfold FindSingleNumber_precond at h_precond
  rcases h_precond with ⟨_hall, huniq⟩
  by_contra hne
  let p : Int → Prop := fun x => nums.count x = 1
  have haF : a ∈ nums.filter p :=
    List.mem_filter.2 ⟨ha, by simp [p, hca]⟩
  have hbF : b ∈ nums.filter p :=
    List.mem_filter.2 ⟨hb, by simp [p, hcb]⟩
  have two_le {α : Type} {l : List α} {a b : α}
      (ha : a ∈ l) (hb : b ∈ l) (hne : a ≠ b) :
      2 ≤ l.length := by
    induction l with
    | nil => cases ha
    | cons c tl ih =>
        have ha' : a = c ∨ a ∈ tl := by simpa using ha
        have hb' : b = c ∨ b ∈ tl := by simpa using hb
        cases ha' with
        | inl hac => cases hb' with
          | inl hbc =>
              exact (hne (by simpa [hac] using hbc.symm)).elim
          | inr hb_tl =>
              exact Nat.succ_le_succ
                (Nat.succ_le_of_lt (List.length_pos_of_mem hb_tl))
        | inr ha_tl => cases hb' with
          | inl hbc =>
              exact Nat.succ_le_succ
                (Nat.succ_le_of_lt (List.length_pos_of_mem ha_tl))
          | inr hb_tl =>
              exact le_trans (ih ha_tl hb_tl) (Nat.le_succ tl.length)
  have hlen_ge : 2 ≤ (nums.filter p).length :=
    two_le haF hbF hne
  have count_map_eq_aux : ∀ l : List Int,
      (l.map (fun x => nums.count x)).count 1 =
        (l.filter (fun a => nums.count a = 1)).length := by
    intro l; induction l with
    | nil => simp
    | cons a l ih =>
        by_cases h : nums.count a = 1 <;> simp [p, h, ih]
  have hlen_eq_one : (nums.filter p).length = 1 := by
    simp only [p]
    exact (count_map_eq_aux nums).symm.trans huniq
  exact Nat.not_succ_le_self 1
    (by simpa [hlen_eq_one] using hlen_ge)
\end{leancode-gray}

\begin{leancode-gray}
-- Lemma 6: non-unique elements have count = 2
lemma count_eq_two_of_ne_of_count_eq_one
    (nums : List Int)
    (h_precond : FindSingleNumber_precond nums)
    (r x : Int) (hx : x ∈ nums)
    (hcr : nums.count r = 1) (hxr : x ≠ r) :
    nums.count x = 2 := by
  classical
  unfold FindSingleNumber_precond at h_precond
  rcases h_precond with ⟨hall, huniq⟩
  have precond_counts (x' : Int) (hx' : x' ∈ nums) :
      nums.count x' = 1 ∨ nums.count x' = 2 := by
    have hall' : ∀ n ∈ (nums.map fun y => nums.count y),
        (n = 1 ∨ n = 2) := by
      simpa [List.all_eq_true] using hall
    exact hall' (nums.count x')
      (List.mem_map.mpr ⟨x', hx', rfl⟩)
  have h_precond' : FindSingleNumber_precond nums := by
    unfold FindSingleNumber_precond; exact ⟨hall, huniq⟩
  have mem_of_one (a : Int) (h : nums.count a = 1) :
      a ∈ nums := by
    by_contra hmem
    have : nums.count a = 0 := by
      simp [List.count_eq_zero, hmem]
    exact Nat.zero_ne_one (by simpa [this] using h)
  cases precond_counts x hx with
  | inl h1 =>
      exact (hxr (precond_unique_count_one nums h_precond'
        hx (mem_of_one r hcr) h1 hcr)).elim
  | inr h2 => exact h2
\end{leancode-gray}

\begin{leancode-gray}
-- Main theorem: combining all lemmas
theorem FindSingleNumber_spec_satisfied
    (nums : List Int)
    (h_precond : FindSingleNumber_precond nums) :
    FindSingleNumber_postcond nums
      (FindSingleNumber nums h_precond) h_precond := by
  classical
  constructor
  · by_contra h0
    have hnil : nums = [] := by
      simpa using List.length_eq_zero_iff.mp
        (Nat.eq_zero_of_not_pos h0)
    subst hnil
    simp [FindSingleNumber_precond] at h_precond
  constructor
  · simpa [length_filterlist_eq_count] using
      FindSingleNumber_result_characterization nums h_precond
  · intro x hx
    by_cases h : x = FindSingleNumber nums h_precond
    · exact Or.inl h
    · have hcount :=
        count_eq_two_of_ne_of_count_eq_one nums h_precond
          (FindSingleNumber nums h_precond) x hx
          (FindSingleNumber_result_characterization
            nums h_precond) h
      exact Or.inr
        (by simp [length_filterlist_eq_count, hcount])
\end{leancode-gray}

\section{Algorithm Pseudocode}
\label{app:algorithms}

We provide the full pseudocode for the two core procedures of our framework.
Algorithm~\ref{alg:hybrid-rl} details the hybrid reinforcement learning loop that jointly optimizes decomposition via GRPO and completion via supervised replay, while progressively expanding the problem set with newly decomposed lemmas.
Algorithm~\ref{alg:inference} describes the two-stage hierarchical inference pipeline, where the policy first recursively decomposes the verification goal and then attempts to prove each remaining leaf lemma with iterative refinement.

\begin{algorithm}[h]
\caption{Hybrid RL with Online Lemma Collection}
\label{alg:hybrid-rl}
\begin{algorithmic}[1]
\REQUIRE Initial problem set $\mathcal{P}_0$, policy $\pi$, loss coefficient $\lambda$
\FOR{iterations $t = 1, 2, \dots$}
    \STATE Sample problem $G \sim \mathcal{P}_{t-1}$.
    \STATE Roll out $\pi$ to produce decomposed lemmas $\mathcal{L}_t := \{L_1, \dots, L_k\}$.
    \STATE Compute reward $S = r(L_1, \dots, L_k; G) \cdot v(L_1, \dots, L_k; G)$ and GRPO objective $\mathcal{J}_{\text{GRPO}}$.
    \STATE Attempt completion of $\mathcal{L}_t$: first roll out $\pi$; if $\pi$ fails within $m$ attempts, fall back to an off-the-shelf model to produce $\mathcal{T}_t := \{\tau_1, \dots, \tau_k\}$.
    \STATE Compute SFT objective $\mathcal{J}_{\text{SFT}}$ on completion trajectories $(\mathcal{L}_t, \mathcal{T}_t)$.
    \STATE Update $\pi$ with hybrid objective $\mathcal{J} = \mathcal{J}_{\text{GRPO}} + \lambda \cdot \mathcal{J}_{\text{SFT}}$.
    \STATE Augment problem set: $\mathcal{P}_t \leftarrow \mathcal{P}_{t-1} \cup \mathcal{L}_t$.
\ENDFOR
\end{algorithmic}
\end{algorithm}

\begin{algorithm}[h]
\caption{Hierarchical Inference (Single Run)}
\label{alg:inference}
\begin{algorithmic}[1]
\REQUIRE Theorem $G$, policy $\pi$
\STATE \textbf{// Stage 1: Lemma Decomposition}
\STATE Initialize open goal set $\mathcal{O} \leftarrow \{G\}$.
\FOR{decomposition iterations $i = 1, \dots, 128$}
    \STATE Select the goal $g \in \mathcal{O}$ with the largest operator footprint.
    \STATE Roll out $\pi$ to decompose $g$ into sub-lemmas $\{L_1, \dots, L_k\}$.
    \STATE Verify proof reconstruction and quickcheck each $L_i$; if either check fails, discard and retry.
    \STATE If the update would exceed 32 open goals, stop decomposition.
    \STATE Update $\mathcal{O}$: remove $g$, add $\{L_1, \dots, L_k\}$; compute operator footprints.
\ENDFOR
\STATE \textbf{// Stage 2: Lemma Completion}
\STATE Collect remaining leaf lemmas $\mathcal{L} = \{L_1, \dots, L_m\}$ from $\mathcal{O}$.
\FOR{completion iterations $j = 1, \dots, 128$}
    \STATE Roll out $\pi$ to generate candidate proofs for each unclosed $L_i \in \mathcal{L}$.
    \STATE Refine proofs based on Lean compiler feedback.
    \STATE \textbf{if} all lemmas closed \textbf{then} \textbf{return} success.
\ENDFOR
\STATE \RETURN failure.
\end{algorithmic}
\end{algorithm}

\section{Implementation Details}
\label{app:implementation}

\paragraph{Custom Lean Tactics.}
We implement two custom Lean tactics used by the decomposition scoring mechanism.
\texttt{operatorcount} traverses the goal's abstract syntax tree and returns the total number of logical operators (connectives, quantifiers) and domain-specific program operators (arithmetic, bitwise, data-structure constructors), excluding variable references and type annotations.
\texttt{quickcheck} extends the \texttt{Plausible} library~\citep{lean2024plausible}: given a universally quantified lemma, it randomly samples concrete inputs (up to 1000 trials) and evaluates the lemma via native execution, reporting a counterexample if one is found.
Both tactics are registered as Lean meta-programs and can be invoked at any proof state during decomposition verification.

\paragraph{Data Curation.}
We collect 5K unique problems from LeetCode~\citep{xia2025leetcodedataset} and OpenCodeInstruct~\citep{ahmad2025opencodeinstruct}.
Each problem is automatically formalized into Lean using GPT-5.2.
The formalizations undergo iterative refinement until they pass both syntactic validation and quickcheck filtering, ensuring empirical consistency.
To prevent data contamination, we remove all problems whose title or description has an exact match with any problem in the Verina, Clever, or AlgoVeri evaluation benchmarks.
We additionally audit the resulting split using longest-shared-substring and embedding-similarity checks on boilerplate-stripped Lean implementations; Appendix~\ref{app:decontamination} reports the thresholds and flagged pairs.

\paragraph{Trajectory Filtering.}
We apply different filtering criteria for the two trajectory types.
For decomposition trajectories, we retain only those that are constructively justified (the proposed lemmas provably entail the parent theorem) and achieve a positive structural-reduction score ($S>0$).
For completion trajectories, we retain only those whose final proof is accepted by the Lean kernel.

\paragraph{Supervised Fine-Tuning.}
We fine-tune Qwen3-8B~\citep{yang2025qwen3} using LLaMA-Factory~\citep{zheng2024llamafactory} for three epochs on a mixture of decomposition and proof completion trajectories.
Training uses a learning rate of $1\times10^{-5}$ with a cosine schedule and a warmup ratio of $0.1$.
The maximum context length is $10{,}240$ tokens with sequence packing enabled.
We use a batch size of $128$ with sequence packing.

\paragraph{Reinforcement Learning.}
We apply GRPO~\citep{shao2024deepseekmath} using the \texttt{verl} framework~\citep{sheng2024hybridflow} for 100 training steps across 16 GPUs.
The learning rate is $5\times10^{-6}$ with cosine decay, with a clip ratio of $[0.2,\,0.28]$ and a sampling temperature of $0.9$.
We use a batch size of 64 prompts with $n{=}8$ parallel generations per prompt and a mini-batch size of 256 for policy update.
Rewards are computed by verifying proof attempts and decomposition quality as described.
To ensure informative gradient signals, we filter out rollout groups whose mean reward is $0$ or $1$.
We additionally apply an auxiliary supervised fine-tuning loss on proof completion trajectories with coefficient $\lambda{=}0.08$ and refresh the online proof-state buffer at a sampling ratio of $\frac{1}{4}$ per step.

\paragraph{Prompting Format.}
For decomposition, the prompt presents the current Lean file with the target theorem highlighted, and instructs the model to produce a \texttt{<reasoning>} block followed by helper lemmas (each with \texttt{sorry} placeholders) and a main theorem whose proof invokes these lemmas without \texttt{sorry}.
For completion, the prompt shows the current goal state and compilation status, and the model responds with a SEARCH/REPLACE diff that modifies only the proof body while preserving theorem signatures and imports.
We use the same prompting format for both training and inference.
The full prompt templates are shown in Figure~\ref{fig:prompt-decompose} and Figure~\ref{fig:prompt-complete}.

\begin{figure*}[t]
\small
\begin{tcolorbox}[colback=codebg, colframe=codeframe, colbacktitle=codeframe!70!black, boxrule=0.5pt, arc=2pt, left=4pt, right=4pt, top=4pt, bottom=4pt, title={\small\textbf{Decomposition Prompt Template}}, fonttitle=\small\bfseries]
\textbf{System:} You are an expert in Lean 4 theorem proving. Your task is to decompose a theorem into smaller, provable lemmas.

\textbf{Think step by step.} Before writing code, first analyze the theorem structure and explain your decomposition strategy in a \texttt{<reasoning>} block.

\textbf{Decomposition Strategies:}
\begin{itemize}[leftmargin=*, nosep]
\item \textbf{Structural Decomposition} --- Match goal structure: split $P \wedge Q$ into separate lemmas; prove one side of $P \vee Q$; find witness for $\exists x, P(x)$.
\item \textbf{Induction Decomposition} --- For recursive structures: base case + inductive step as separate lemmas.
\item \textbf{Rewrite Chain} --- For equational goals $A = D$: break into $A = B$, $B = C$, $C = D$.
\item \textbf{Case Analysis} --- Split by \texttt{Decidable}, \texttt{match}, or \texttt{if-then-else}; each branch becomes a separate lemma.
\end{itemize}

\textbf{Output Requirements:}
(1) Wrap analysis in \texttt{<reasoning>...</reasoning>} tags before any code;
(2) Helper lemmas use \texttt{lemma} keyword with \texttt{sorry} body;
(3) Main theorem uses \texttt{theorem} keyword and proves using the lemmas (no \texttt{sorry}).

\vspace{4pt}
\textbf{User:} Decompose the theorem \texttt{\{theorem\_name\}}. \\ \texttt{\{formal\_problem\}}
\end{tcolorbox}
\caption{Prompt template for the decomposition stage. The model receives the target theorem name and the full Lean formalization, and produces a reasoning trace followed by helper lemmas and a proof of the main theorem.}
\label{fig:prompt-decompose}
\end{figure*}

\begin{figure*}[t]
\small
\begin{tcolorbox}[colback=codebg, colframe=codeframe, colbacktitle=codeframe!70!black, boxrule=0.5pt, arc=2pt, left=4pt, right=4pt, top=4pt, bottom=4pt, title={\small\textbf{Completion Prompt Template}}, fonttitle=\small\bfseries]
\textbf{System:} Lean 4 proof assistant. Complete the proof by filling in the \texttt{sorry} placeholder with valid tactics.

\textbf{Important:} Output reasoning in a \texttt{reasoning} block before any diff blocks.

\textbf{Recommended tactics} (try in order): \texttt{grind} (powerful automation), \texttt{aesop} (general automation), \texttt{simp\_all} (aggressive simplification), \texttt{omega} (linear arithmetic), \texttt{by\_cases} (case split), \texttt{induction} (induction with auto finish).

\textbf{Output format:} SEARCH/REPLACE diff blocks.

\vspace{4pt}
\textbf{User:} Complete the current proof goal.

\texttt{\{code\}}

Goal (Line \texttt{\{line\}}): \texttt{\{goal\}}

After the reasoning block, output diff blocks only.
\end{tcolorbox}
\caption{Prompt template for the completion stage. The model receives the current code, the target goal state, and responds with SEARCH/REPLACE diffs that fill in proof bodies.}
\label{fig:prompt-complete}
\end{figure*}

\paragraph{Inference Configuration.}
Each problem is allocated a wall-clock budget of 30 minutes.
The decomposition stage runs up to 128 iterations with a maximum of 32 lemmas per problem.
The completion stage runs up to 128 iterations per lemma, with a two-phase strategy: (1)~attempt tactic-based proving when the code compiles, and (2)~fix compilation errors if present.
We deploy a distributed Lean verification service built on Ray~\citep{moritz2018ray}, which manages a pool of Lean worker processes across multiple nodes and supports up to 512 concurrent verification requests with a 300-second timeout per check.

\section{Additional Results}
\label{sec:appendix}

\subsection{Off-the-Shelf Open-Model Instantiation}
\label{app:open-model}

\begin{table}[h]
    \centering
    \caption{Verina results (\%) when instantiating the framework with off-the-shelf models without fine-tuning.}
    \label{tab:open-model}
    \setlength{\tabcolsep}{4pt}
    \renewcommand{\arraystretch}{1.15}
    \begin{tabular}{llr}
    \toprule
    Decomposition & Completion & Verina \\
    \midrule
    --- & GPT-OSS-120B & 20.1 \\
    --- & Gemini-3-Flash & 19.6 \\
    GPT-OSS-120B & GPT-OSS-120B & 44.9 \\
    GPT-5.2-Pro & Gemini-3-Flash & 54.4 \\
    \bottomrule
    \end{tabular}
\end{table}

Using GPT-OSS-120B for both stages raises its success rate from 20.1\% in whole-proof generation to 44.9\% with hierarchical search.
This result shows that the structural benefit of decomposition is not restricted to the frontier models used for data collection; it does not, however, substitute for a fully frontier-free training run.

\subsection{Training--Evaluation Decontamination Audits}
\label{app:decontamination}

We supplement title/description exact matching with two audits on boilerplate-stripped Lean implementations.
First, following the ExactSubstr-style audit of \citet{lee2022deduplicating}, we compare each benchmark implementation with all 4,277 retained training implementations.

\begin{table}[h]
    \centering
    \caption{Benchmark tasks whose implementation shares a contiguous substring with a training implementation.}
    \label{tab:substring-audit}
    \setlength{\tabcolsep}{4pt}
    \renewcommand{\arraystretch}{1.15}
    \begin{tabular}{lrrrr}
    \toprule
    Threshold & Verina & Clever & AlgoVeri & Total \\
    \midrule
    $\geq 50$ chars  & 7 & 5 & 1 & 13/427 \\
    $\geq 100$ chars & 3 & 0 & 0 & 3/427 \\
    $\geq 200$ chars & 0 & 0 & 0 & 0/427 \\
    \bottomrule
    \end{tabular}
\end{table}

The Clever and AlgoVeri hits at 50 characters are generic Lean fragments rather than shared problem content; neither benchmark has a hit at 100 characters, and no benchmark task has a hit at 200 characters.
Second, we embed each stripped reference implementation with \texttt{text-embedding-3-large} and flag nearest training neighbors with cosine similarity at least 0.95.
Only 2/427 tasks (0.47\%), both from Verina, are flagged: \texttt{verina\_advanced\_42}/\texttt{maxProfit} (0.997) and \texttt{verina\_advanced\_18}/\texttt{isArmstrong} (0.964).
Manual inspection finds materially different specifications or implementations in both pairs, so neither admits direct proof transfer.
Together with title/description filtering, these audits find no evidence of proof-level leakage; the residual Verina similarity is consistent with independently formalizing tasks from the same LeetCode source.

\subsection{Decomposition Score Beyond the Evaluation Benchmarks}
\label{app:score-case-study}

We evaluate the operator-footprint reduction ratio $r$ on 22 CSLib and 10 Mathlib theorems with human-written lemma decompositions.

\begin{table}[H]
    \centering
    \caption{Behavior of the operator-footprint reduction ratio on human-written decompositions. The ratio is informative primarily for Hoare-style program-correctness statements, its intended setting.}
    \label{tab:score-case-study}
    \small
    \setlength{\tabcolsep}{4pt}
    \renewcommand{\arraystretch}{1.15}
    \begin{tabular}{@{}p{3.0cm}ccp{8.6cm}@{}}
    \toprule
    Theorem class & Cases & $r>0$ & Representative examples (reduction ratio) \\
    \midrule
    Hoare-style program correctness (CSLib) & 14 & 9 & \texttt{otp\_ciphertextDist\_eq\_uniform} (0.872), \texttt{mergeSort\_sorted} (0.502), \texttt{toNAFinAcc\_language\_eq} (0.492), Shamir privacy (0.154), and \texttt{pred\_correct} for SKI (0.112) \\
    Type-safety theorems (CSLib) & 3 & 0 & STLC/Fsub preservation and progress statements \\
    Structural and categorical theorems (CSLib) & 5 & 0 & LTS composition, Turing-machine composition, $\omega$-regular complement, CLL $\eta$-expansion, and infinite Ramsey \\
    Pure mathematics (Mathlib) & 10 & 0 & Lagrange, Beatty, Hensel, Siegel, Kronecker, Shortlex, Lucas--Lehmer, and related theorems \\
    \bottomrule
    \end{tabular}
\end{table}

Nine of the 14 Hoare-style cases receive a positive reduction ratio, while all abstract type-safety, structural, categorical, and pure-mathematics cases receive zero.
This behavior matches the score's intended asymmetry: program postconditions often unfold into many program operators, whereas useful helper lemmas isolate simpler sub-properties.
The result also delineates the metric's scope---operator-footprint reduction is not a general-purpose measure of mathematical decomposition quality.

\subsection{Inference Budgets and Comparison Scope}
\label{app:budget-comparison}

Our system allocates 30 minutes per problem, runs up to 128 decomposition iterations with at most 32 open lemmas, and then runs up to 128 completion iterations per lemma; pass@$k$ launches $k$ independent runs.
The reported baselines use their default decoding configurations at pass@128, and the baseline scaling curves in Figure~\ref{fig:pass-at-k} extend BFS-Prover-V2 and G\"{o}edel-Prover-V2 to pass@1024.
Because closed models expose token/API accounting while open provers use heterogeneous model sizes and decoding implementations, these evaluations are not compute-matched.
Accordingly, our comparisons concern verified success under the stated budgets, not equal-cost efficiency or an effect attributable to parameter count.

\subsection{Proof Statistics}
\label{app:proof-stats}

\begin{table}[h]
    \centering
    \caption{Statistics of proved problems: mean and standard deviation of lemma count and proof length (lines of code).}
    \label{tab:lemma-proof-stats}
    \begin{threeparttable}
    \setlength{\tabcolsep}{6pt}
    \renewcommand{\arraystretch}{1.2}
    \begin{tabular}{l|>{\centering\arraybackslash}p{1.6cm}|>{\centering\arraybackslash}p{1.6cm}|>{\centering\arraybackslash}p{1.6cm}|>{\centering\arraybackslash}p{1.6cm}}
    \toprule
    Statistics  & {Verina} & {Clever} & {AlgoVeri} & {Total} \\
    \midrule
    Lemma count (mean)  & 17.02 & 12.13 & 8.48 & 14.38 \\
    Lemma count (std)    & 11.79 & 10.57 & 10.97 & 11.97 \\
    \midrule
    Proof length (mean) & 167.11 & 137.78 & 129.87 & 154.28 \\
    Proof length (std)  & 108.50 & 97.79 & 83.03 & 103.11 \\
    \bottomrule
    \end{tabular}
    \end{threeparttable}
\end{table}

Table~\ref{tab:lemma-proof-stats} reports the mean and standard deviation of lemma count and proof length (in lines of code) for problems successfully proved by our framework across the three benchmarks.

\subsection{Residual Operator-Footprint Ratio}
\label{app:reduction-rate}

\begin{figure}[h]
    \centering
    \includegraphics[width=\linewidth]{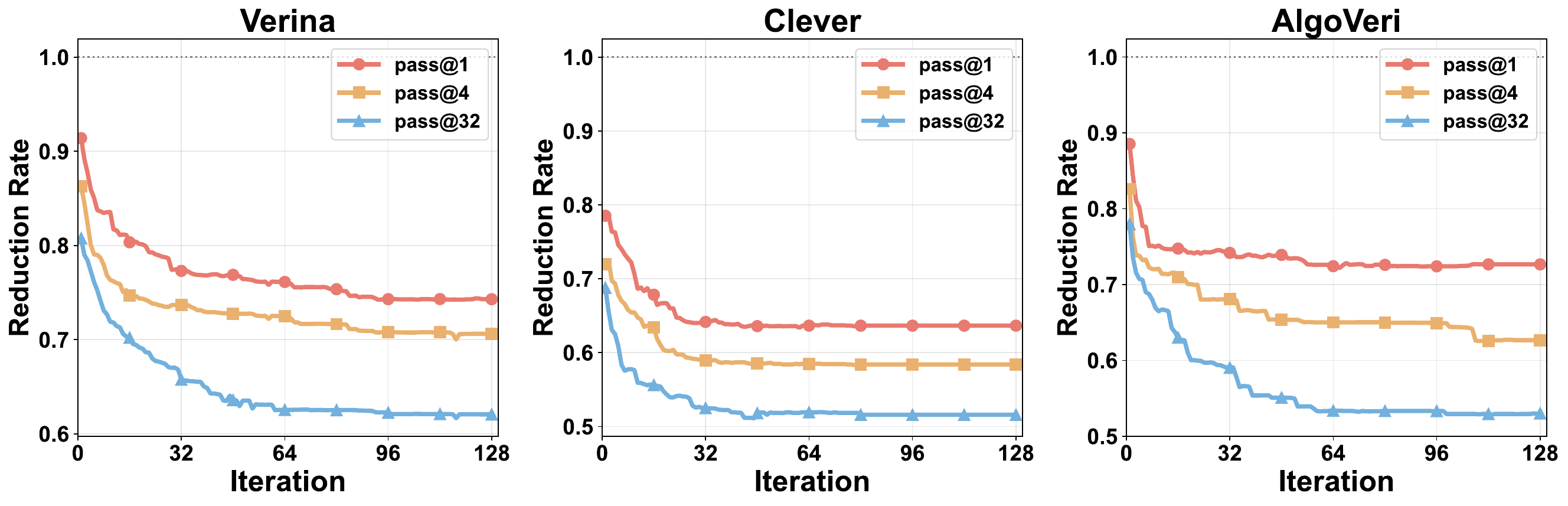}
    \caption{Residual operator-footprint ratio $\bar d/d(G)$ vs.\ iterations across three benchmarks. Lower values indicate more aggressive goal simplification and therefore a higher reduction ratio $r$.}
    \label{fig:reduction-rate}
\end{figure}

Figure~\ref{fig:reduction-rate} shows how the residual operator-footprint ratio evolves as iterations progress under different pass@$k$ budgets. The ratio decreases steadily, indicating that the framework identifies increasingly effective goal simplifications with additional compute.

\subsection{Baseline Inference Scaling Curves}
\label{app:pass-at-k}

\begin{figure}[h]
    \centering
    \includegraphics[width=\linewidth]{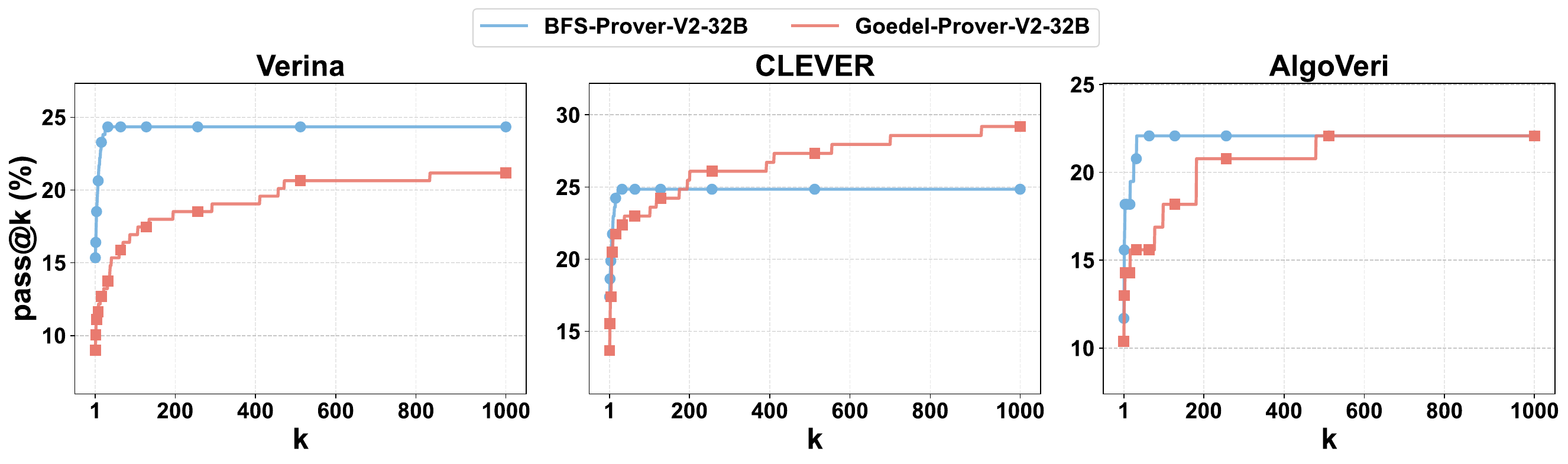}
    \caption{Inference scaling curves (pass@k) of two baseline verifiers, BFS-Prover-V2-32B and G\"{o}edel-Prover-V2-32B, on the three evaluation benchmarks. The x-axis denotes the number of sampled proof attempts $k$, and the y-axis shows the percentage of problems solved.}
    \label{fig:pass-at-k}
\end{figure}

To contextualize the inference-time scaling behavior of our framework, we examine the pass@$k$ curves of the two strongest baseline verifiers.
As shown in Figure~\ref{fig:pass-at-k}, both BFS-Prover-V2-32B and G\"{o}edel-Prover-V2-32B exhibit diminishing returns as $k$ increases: performance plateaus well before $k{=}1024$ on all three benchmarks, suggesting that simply scaling the number of independent proof attempts provides limited gains for whole-proof generation approaches.
In contrast, our hierarchical framework continues to improve with additional compute (Section~\ref{subsec:rq2}), highlighting the advantage of structured search over brute-force sampling.

\subsection{Decomposition Score Correlation on Training Data}
\label{app:train-correlation}

\begin{figure}[!htbp]
    \centering
    \includegraphics[width=0.93\linewidth]{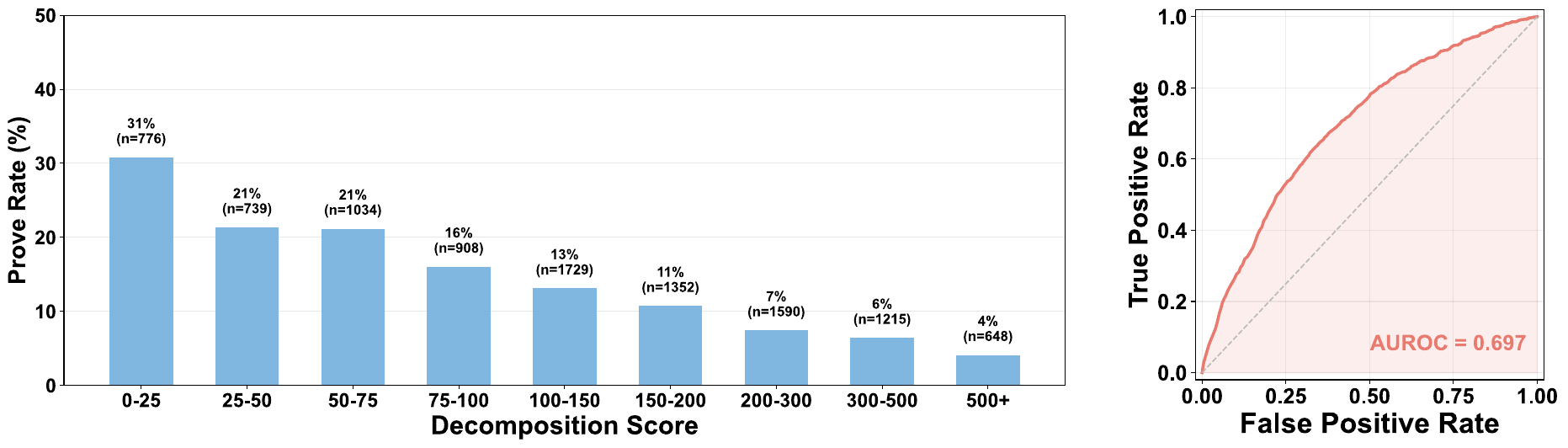}
    \caption{Unnormalized decomposition score vs.\ prove rate on training data. As on Verina, lower raw scores are associated with higher prove rates.}
    \label{fig:train-correlation}
\end{figure}

In Section~\ref{subsec:rq3}, we show that the unnormalized decomposition score is a strong inverse predictor of downstream provability on Verina (AUROC = 0.903).
We repeat this analysis on the training data to determine whether the association is unique to that evaluation set.
As shown in Figure~\ref{fig:train-correlation}, problems with lower raw scores are also more likely to be proved on the training distribution.
This in-distribution result shows that the association is not unique to Verina, but it does not by itself establish out-of-distribution generalization.

\subsection{Effect of Training on Decomposition Quality}
\label{app:score-distribution}

Across Verina, Clever, and AlgoVeri, the mean raw score falls from 282 to 185, 437 to 329, and 271 to 196, respectively; lower raw scores correspond to structurally simpler decompositions and higher normalized $S$.

\begin{figure}[!htbp]
    \centering
    \includegraphics[width=0.93\linewidth]{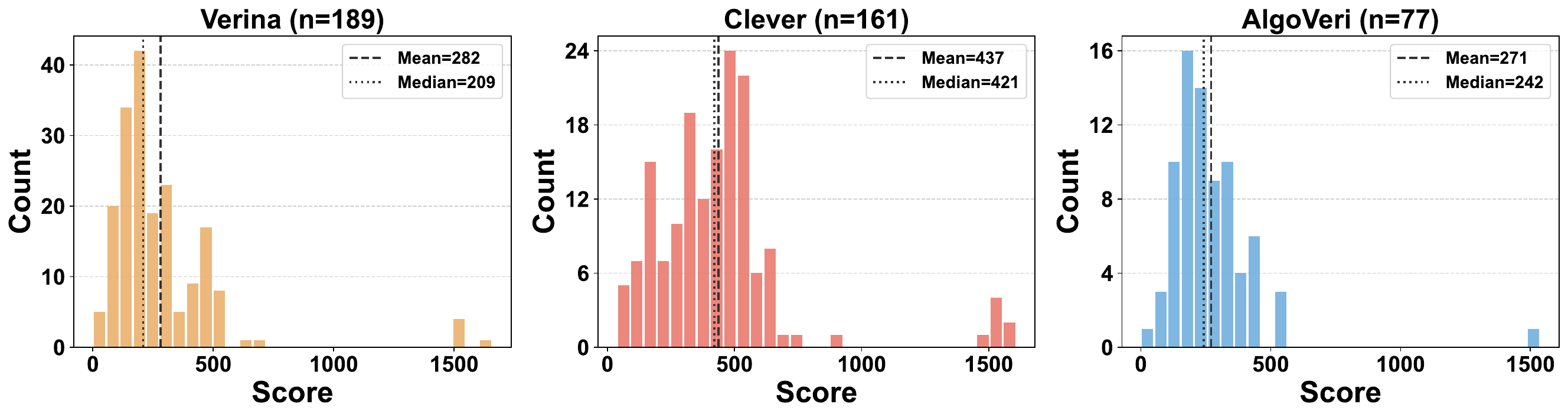}
    \caption{Unnormalized decomposition-score distribution before training across benchmarks. Lower values are better.}
    \label{fig:score-dist-before}
\end{figure}

\begin{figure}[!htbp]
    \centering
    \includegraphics[width=0.93\linewidth]{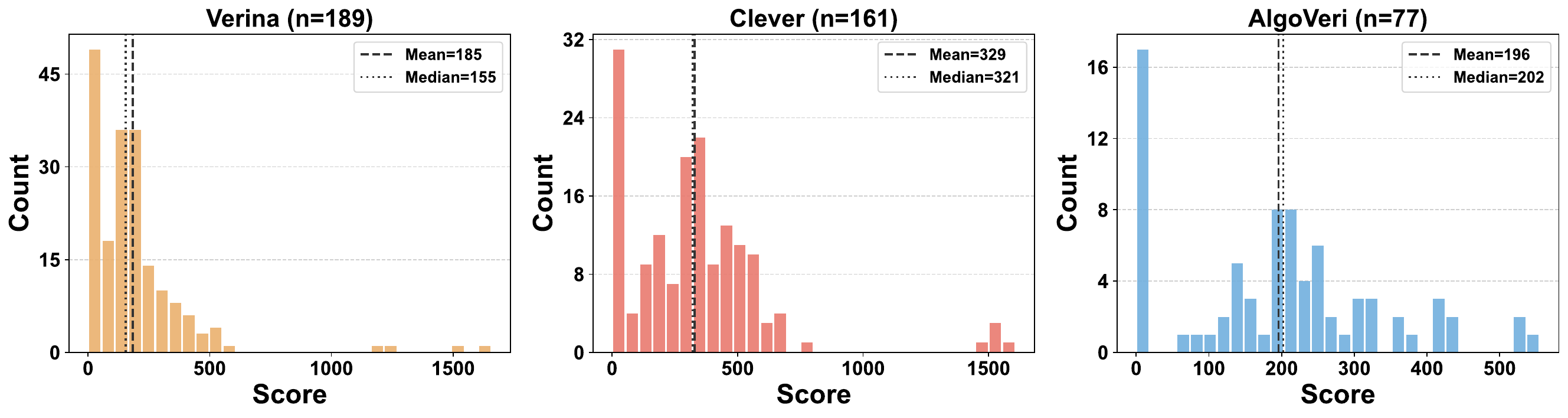}
    \caption{Unnormalized decomposition-score distribution after training across benchmarks. Lower values are better.}
    \label{fig:score-dist-after}
\end{figure}

\end{document}